\documentclass[11pt]{article}

\usepackage[utf8]{inputenc} 
\usepackage{comment}

\usepackage[numbers]{natbib}
\usepackage{hyperref}
\usepackage{geometry} 
\usepackage{graphicx} 
\usepackage{booktabs}
\usepackage{array} 
\usepackage{paralist} 
\usepackage{verbatim}
\usepackage{bm} 
\usepackage{subfig} 
\usepackage{amsmath}
\usepackage{amssymb}

\usepackage{xcolor}
\usepackage{caption}

\usepackage{fancyhdr} 

\usepackage{sectsty}
\allsectionsfont{\sffamily\mdseries\upshape}
\usepackage[nottoc,notlof,notlot]{tocbibind} \usepackage[titles,subfigure]{tocloft}

\numberwithin{equation}{section}

\title{Revisiting the Bohr Model of the Atom through Brownian Motion of the Electron}
\author{Vasil Yordanov \\
	v.yordanov@phys.uni-sofia.bg \\
	\textit{Faculty of Physics, Sofia University,}\\
	\textit{5 James Bourchier blvd., 1164 Sofia, Bulgaria}
}

\begin{document}
\maketitle
\abstract
We revisit the Bohr model of the atom through Brownian motion of the
electron and the principles of stochastic optimal control. The
electron is assumed to have a definite but random position, represented by
a single real-valued stochastic process in physical space whose probability
density obeys the Fokker--Planck equation. Because Brownian paths are not
differentiable, the
process carries two mean drifts, one for each direction of time. We treat
the forward drift as the control field, while the backward drift is fixed
by the density of the same process. The running cost of the control problem
combines the two drifts into a time-symmetric kinetic term, and through the
backward drift it inherits a dependence on the density, so the value
becomes a functional on density space. Bellman's dynamic-programming
principle requires the control to minimize the expected action from every
intermediate time and density onward. The drift therefore emerges as a feedback law on position and
density, rather than from the global stationarity of a stochastic action.
The resulting law-dependent HJB--Fokker--Planck system reduces to the
Schr\"odinger equation once the standard single-valuedness condition is
imposed on the reconstructed phase.

For stationary hydrogen states the theory yields explicit drift fields in
spherical coordinates and reproduces the standard radial and angular
kinetic-energy averages of the quantum operator formalism. Direct
trajectory-level simulations of the electron's stochastic motion around the
nucleus show the coordinate distributions converging to the Born marginals
and the time-averaged energies reproducing the quantum expectation values.
For the \(2p\) eigenstates with magnetic quantum number \(m=\pm1\), a
phase-driven azimuthal drift makes the simulated trajectories circulate at
the analytically predicted rate, and the angular momentum accumulated from
the raw trajectory increments converges to exactly \(L_z=m\hbar\). The
angular-momentum quantization postulated in the Bohr model thus reappears
as a property of the simulated stochastic motion.

\section{Introduction}
\label{sec:introduction}

The hydrogen atom was historically the simplest system in which the conflict
between classical orbital mechanics and atomic spectra became unavoidable.
Thomson's model~\cite{Thomson1904} and Rutherford's nuclear atom
\cite{Rutherford1911} established the basic picture of electrons bound to a
central charge, while Bohr's model~\cite{Bohr1913} explained the hydrogen
spectrum by imposing quantized orbits on an otherwise classical dynamics.
Schr\"odinger's wave mechanics~\cite{Schrodinger1926} replaced these orbits by
a wave equation and a probabilistic description. The present paper revisits this
transition from a different direction: it asks whether the stationary quantum
properties of hydrogen can be represented by a single real stochastic motion of
the electron in physical space. We develop such a representation by combining
Nelson's forward--backward stochastic kinematics~\cite{Nelson1966} with
Bellman's dynamic-programming principle~\cite{bellman1954} in a control problem
posed on the space of probability densities. This construction determines the
forward drift of the electron's stochastic motion as an optimal feedback field.

In Nelson's framework,  the electron's motion is modeled as a diffusion with
coefficient \(D=\hbar/(2m)\).
This diffusion admits forward \(\mathbf b_{+}\) and backward \(\mathbf b_{-}\) drift fields, associated with its two time directions~\cite{Nelson1966}.
The forward--backward kinematics is essential: the difference between the two drifts is fixed by the probability density of the particle's position through the osmotic velocity \(\mathbf u=\frac12 (\mathbf b_{+}-\mathbf b_{-})=D\,\nabla\!\ln P\),  while their mean defines the current velocity \(\mathbf v=\frac12 (\mathbf b_{+}+\mathbf b_{-})\).

Nelson's theory was subsequently reformulated and extended through stochastic
variational methods. Because the present construction is posed on density
space, a point of particular relevance is how probability-density evolution
enters these variational approaches. Yasue~\cite{Yasue1981} proposed a
time-symmetric stochastic
variational formulation of quantum mechanics with the symmetric kinetic
Lagrangian \(\tfrac{1}{2}m(\mathbf v^{2}+\mathbf u^{2})-V\). The stochastic
Euler--Lagrange equation is obtained without imposing the Fokker--Planck
equation as a variational constraint. The Fokker--Planck equation nevertheless
enters Yasue's treatment of the density evolution and the equivalence with the
Schr\"odinger initial-value problem. Guerra and Morato~\cite{Guerra1983}
introduced a stochastic action principle with the alternative time-symmetric
Lagrangian \(\tfrac{1}{2}m(\mathbf v^{2}-\mathbf u^{2})-V\), which recovers
Nelson's mechanics and yields the Madelung equations~\cite{Madelung1927}. In
their construction, the controlled stochastic differential equation gives an
explicit evolution equation for the one-time density, while the transition
probability density satisfies the forward Fokker--Planck equation. A variation
of the control drift induces corresponding variations of both densities.
Pavon developed related variational formulations --- in both
hydrodynamic~\cite{Pavon1995} and particle~\cite{Pavon1996} forms --- in which
the Madelung equations arise as Hamilton--Jacobi-like equations of a stochastic
variational problem.

More recently, Yang~\cite{Yang2021_Jianhao} proposed a stochastic variational
formulation that supplements the least-action principle with constraints based
on relative entropy and Fisher information, treating each forward and backward
path configuration as a variational degree of freedom. The Fokker--Planck
equations are not imposed as variational inputs. A relation obtained from the
variation is combined with the forward--backward drift relation, first to
obtain the continuity equation and then to rewrite it as the corresponding
forward and backward Fokker--Planck equations.

The present formulation retains Nelson's forward--backward description of the
electron as a single stochastic process with drifts \(\mathbf b_+\) and
\(\mathbf b_-\). It adds an optimal-control structure by treating \(\mathbf b_+\)
as the control field, while \(\mathbf b_-\) is not an independent control: for
the same process, it is fixed by \(\mathbf b_+\) and the probability density
\(P\) through the forward--backward time-reversal relation. We choose the
bilinear Lagrangian
\(L=\tfrac12m\,\mathbf b_+\!\cdot\!\mathbf b_- -V
=\tfrac12m(\mathbf v^2-\mathbf u^2)-V\) as the running cost.
Expressing \(\mathbf b_-\) through \(\mathbf b_+\) and \(P\) makes this cost
depend on the probability law represented by \(P\). The control problem is
therefore law-dependent and is posed on density space, with \(P\) as the state
variable. The Bellman value is consequently a functional \(\mathcal V(t,P)\)
on density space. Bellman's principle yields the functional HJB equation, and
minimization over the forward-drift control field determines the optimal
feedback drift. For each admissible forward drift, the controlled stochastic
differential equation induces the evolution of \(P\) through the forward
Fokker--Planck (FP) equation. This density evolution is kinematic and does not
follow from Bellman's principle; accordingly, the chosen time-symmetric cost
\(L\) neither avoids nor replaces the FP equation. The FP and functional HJB
equations together form the coupled HJB--FP system governing the density-space
control problem.

The present optimal-control formulation differs from the
Guerra--Morato--Pavon variational formulations in how optimality is imposed. In
those formulations, the dynamics follow from stationarity of a stochastic
action under admissible variations over a prescribed time interval. Here,
Bellman optimality is recursive: the continuation cost is minimized from every
intermediate time and probability density.

This law-dependent density-space structure is therefore close
to McKean--Vlasov control and dynamic programming on spaces of probability
measures~\cite{Vlasov1961,McKean1966,CarmonaDelarue2018,Pham2017,
BensoussanFrehseYam2013,DjetePossamaiTan2022,GangboMayorgaSwiech2021}.
Unlike standard mean-field applications, however, \(P\) is here the
one-particle density of a single real diffusion, not an empirical density of
interacting particles; the law-dependence comes instead from the
forward--backward consistency of the same diffusion. The
density-space Bellman formulation presented here leads to a local
HJB--Fokker--Planck
system for \(J\) and \(P\). After rewriting this system in terms of the
phase \(S\), one obtains the Madelung equations and, after imposing the
standard single-valuedness condition on the reconstructed
phase~\cite{Wallstrom1989,Wallstrom1994}, the Schr\"odinger equation.

A complementary link between stochastic control and quantum-like dynamics is
provided by the Schr\"odinger bridge problem, in which a control drift steers
a diffusion between prescribed initial and final marginals at minimal
relative entropy~\cite{Schrodinger1931,Leonard2014,
ChenGeorgiouPavon2016,ChenGeorgiouPavon2021}. Its fluid-dynamic formulation
leads to hydrodynamic systems closely related to the Madelung equations, but
there the density enters through the marginal constraints rather than through
the running cost.

A different family of control-theoretic approaches to quantum mechanics uses
complex variables, complex-valued velocities, or imaginary/complex diffusion
coefficients in order to connect stochastic optimal control more directly with
the complex Schr\"odinger equation
\cite{Papiez1982,RosenbrockDing2008,Lindgren2019,Yang2021_Cian_Dong,Yordanov2024}.
Such formulations are useful because the complex phase of the wave function is
built into the stochastic dynamics from the beginning. For the present purpose,
however, they are less direct computationally, because the Born probability is recovered from the statistics of
intersections with the real axis, \(y=0\), rather than from the marginal density
of the full complex process~\cite{Yang2021_Cian_Dong}. The mathematical structure underlying these complex stochastic optimal control formulations is analyzed in our earlier work~\cite{Yordanov2024b}.

After deriving the local law-dependent HJB--Fokker--Planck system, we use the
hydrogen atom as a concrete test case in which the stochastic motion of the
electron can be simulated from the drift field determined by the theory. For
stationary hydrogen states, this drift is available explicitly, allowing
electron trajectories of the same real diffusion to be generated. The
simulations therefore provide a direct way to examine the Born distributions,
the stochastic kinetic-energy averages, and the behavior of the process near
nodal sets where the drift becomes singular.

The contribution of the paper is threefold. First, we formulate the
density-space Bellman problem generated by the forward--backward kinematics
and derive the local law-dependent HJB--Fokker--Planck system, whose Madelung
form recovers the Schr\"odinger equation once the single-valuedness condition
is imposed on the reconstructed phase. Second, for stationary hydrogen states
we obtain closed-form optimal drift fields in spherical coordinates and show
analytically that the stochastic radial and angular kinetic-energy averages
coincide with the corresponding quantum operator expectation values. Third,
we present --- to our knowledge --- the first trajectory-level simulations
of hydrogen eigenstates with nonzero angular momentum, and the first in
which the electron's energies and angular momentum are recovered
quantitatively from the simulated trajectories themselves. The simulations
reproduce the Born marginals for the states considered, and the
time-averaged kinetic, potential, and total energies converge to the
quantum expectation values. The trajectories moreover resolve a
phase-driven azimuthal circulation of the \(2p\) eigenstates with magnetic
quantum number \(m=\pm1\): two states with
identical Born densities but opposite circulation, whose drift carries
exactly \(L_z=m\hbar\) along every trajectory --- a value recovered
numerically from the raw azimuthal increments of the simulated paths --- so
the angular-momentum quantization postulated in the Bohr model
reappears as a deterministic property of the stochastic motion.

The paper is organized as follows. Section~\ref{sec:soc-fp-foundations}
develops the law-dependent stochastic optimal-control formulation and derives
the local HJB--Fokker--Planck system, together with its connection to the
Madelung and Schr\"odinger equations. Section~\ref{sec:spherical-drift-central-potentials}
specializes the stochastic equation of motion to central potentials in spherical
coordinates.
Section~\ref{sec:wavefunction-drift-hydrogen} derives the drift
fields for stationary hydrogen states from the reconstructed wave function.
Sections~\ref{sec:avg_radial_kinetic_energy} and
\ref{sec:avg_angular_kinetic_energy} compare the stochastic radial and angular
kinetic energies with the corresponding operator expressions. The numerical
implementation is described in Section~\ref{sec:hydrogen-simulation}; trajectory
visualizations and coordinate distributions are presented in
Sections~\ref{sec:hydrogen-trajectories-3d} and
\ref{sec:coordinate-distributions}, including the phase-driven azimuthal
circulation of the \(m\neq0\) states
(Section~\ref{subsec:azimuthal-current}). Section~\ref{sec:electron-energies}
compares the simulated energy averages with the analytical values, and
Section~\ref{sec:simulation_convergence_2p0} discusses stability and convergence
near nodal sets.

\section{Law-Dependent HJB--Fokker--Planck Foundations of Quantum Mechanics}
\label{sec:soc-fp-foundations}

Consider a non-relativistic particle of mass \(m\) and position
\(\mathbf x\in\mathbb R^3\), moving in an external potential
\(V(\mathbf x,t)\). In the present formulation the particle is described by a
real-valued Markov diffusion in physical space. Its forward stochastic
differential equation is
\begin{equation}
\label{eq:forward_sde}
dX_t^k
=
b_+^k(\mathbf X_t,t)\,dt
+
\sigma\,dW_t^k,
\qquad k=1,2,3,
\end{equation}
where \(\mathbf b_+\) is the forward drift field, \(\sigma\) is the noise
amplitude, and \(dW_t^k\) are increments of independent standard Wiener
processes.

For an It\^o diffusion of the form~\eqref{eq:forward_sde}, the diffusion
coefficient entering the Fokker--Planck equation is \(D=\sigma^2/2\). Following
Nelson's stochastic mechanics~\cite{Nelson1966}, we choose
\begin{equation}
\label{eq:nelson_diffusion_coefficient}
D
=
\frac{\hbar}{2m}
=
\frac{\sigma^2}{2}.
\end{equation}
This value is motivated by Heisenberg's uncertainty principle, following
Nelson's heuristic argument~\cite{Nelson1985}. For a free
particle of mass \(m\), a position measurement to accuracy \(\Delta x\) imparts
an uncontrolled momentum \(\Delta p\) with \(\Delta p\,\Delta x\geq\hbar/2\),
contributing an additional position shift \(\Delta p\,dt/m\) by time \(t+dt\).
By the arithmetic--geometric-mean inequality, the variance of the observed
displacement around its true value therefore satisfies
\(
\Delta x^{2}+(\Delta p\,dt/m)^{2}
\;\geq\;2(\Delta p\,\Delta x)\,dt/m
\;\geq\;(\hbar/m)\,dt.
\)
The intrinsic mean-square displacement of the diffusion~\eqref{eq:forward_sde}
is \(\mathbb E[d\xi^{2}]=\sigma^{2}\,dt\). Nelson~\cite{Nelson1985} postulates
(the ``background field hypothesis'') that this intrinsic quantity
\emph{saturates} the measurement-uncertainty bound, giving
\(\sigma^{2}=\hbar/m\) and
\(D=\hbar/(2m)\): any larger value would produce intrinsic fluctuations
exceeding the Heisenberg measurement scale and lead to observable deviations
from quantum mechanics, while the exact saturation of the bound is a
postulate rather than a consequence of the argument. The resulting \(1/m\)
scaling makes the diffusion
negligible at macroscopic masses and significant at microscopic ones.

The probability density \(P(\mathbf x,t)\) associated with the stochastic
differential equation~\eqref{eq:forward_sde} evolves according to the
Fokker--Planck (Kolmogorov forward) equation
\begin{equation}
\label{eq:forward_fp}
\partial_t P
=
-\nabla\cdot(\mathbf b_+P)
+
D\Delta P.
\end{equation}
Equivalently,
\begin{equation}
\label{eq:forward_fp_current_form}
\partial_t P
=
-\nabla\cdot
\left[
\mathbf b_+P-D\nabla P
\right],
\end{equation}
so that the forward probability current is
\begin{equation}
\label{eq:probability_current_forward}
\mathbf j
=
\mathbf b_+P-D\nabla P.
\end{equation}

The same diffusion also admits a backward-time description with backward drift
\(\mathbf b_-\)~\cite{Nelson1966,Nelson1985}. The two drifts do not describe
two independent stochastic processes. They describe the same real diffusion
viewed in the two time directions: once the forward drift and the density are
fixed, the backward drift is fixed by the requirement that the same probability
density be represented consistently in both descriptions.

The corresponding
backward Fokker--Planck equation is
\begin{equation}
\label{eq:backward_fp}
\partial_t P
=
-\nabla\cdot(\mathbf b_-P)
-
D\Delta P.
\end{equation}
Subtracting Eq.~\eqref{eq:backward_fp} from Eq.~\eqref{eq:forward_fp} gives
\begin{equation}
\label{eq:drift_difference_divergence}
\nabla\cdot\left[(\mathbf b_+-\mathbf b_-)P\right]
=
2D\Delta P.
\end{equation}
Equation~\eqref{eq:drift_difference_divergence} determines
\((\mathbf b_+-\mathbf b_-)P\) only up to a divergence-free field. The
standard Nelson time reversal of the diffusion~\cite{Nelson1966,Nelson1985}
corresponds to setting this field to zero, which for a normalizable density
is the unique choice compatible with the gradient form of the drifts used
throughout this paper (the optimal drift derived below is a gradient,
Eq.~\eqref{eq:optimal_forward_drift}). With this choice, the relation gives
\begin{equation}
\label{eq:backward_drift_relation}
\mathbf b_-
=
\mathbf b_+-2D\nabla\log P
=
\mathbf b_+
-
\frac{\hbar}{m}\nabla\log P.
\end{equation}
Defining the current velocity by
\begin{equation}
\label{eq:v_definition}
\mathbf v
=
\frac{\mathbf b_++\mathbf b_-}{2},
\end{equation}
the forward and backward drifts are expressed as
\begin{equation}
\label{eq:b_pm_v}
\mathbf b_\pm
=
\mathbf v
\pm \mathbf
u,
\end{equation}
where
\begin{equation}
\label{eq:osmotic_velocity}
\mathbf u
=
\frac{\hbar}{2m}\nabla\log P
\end{equation}
is Nelson's osmotic velocity~\cite{Nelson1966}. Adding
Eqs.~\eqref{eq:forward_fp} and~\eqref{eq:backward_fp} eliminates the diffusion
term and gives the continuity equation
\begin{equation}
\label{eq:continuity_equation}
\partial_t P+\nabla\cdot(P\mathbf v)=0,
\end{equation}
which identifies \(\mathbf v\) as the velocity of probability transport, with
\begin{equation}
\label{eq:current}
\mathbf j=P\mathbf v.
\end{equation}

The forward and backward Fokker--Planck equations are kinematic in origin and
independent of the Bellman minimization introduced below. Only the forward
equation enters the density-space control problem, where it serves as the state
equation. The backward form is used to obtain the same-process time-reversal
relation between \(\mathbf b_-\), \(\mathbf b_+\), and \(P\) in
Eq.~\eqref{eq:backward_drift_relation}.

\subsection{Time-symmetric Lagrangian and the law-dependent running cost}

For the controlled stochastic process considered here, the classical running
cost is \(m\mathbf b_{+}^{2}/2-V\), built from the forward drift alone. This
forward-only cost privileges one time direction. Since the quantum dynamics
to be recovered is time-reversal invariant, it is natural to require the
stochastic kinetic term to treat the forward and backward descriptions
symmetrically. Following the forward--backward kinematics of stochastic
mechanics~\cite{Yasue1981,Guerra1981,Guerra1983,Pavon1995}, we therefore
consider kinetic terms invariant under the interchange
\(\mathbf b_+\leftrightarrow\mathbf b_-\). This symmetry requirement,
however, does not determine a unique stochastic Lagrangian. Within the class
of local scalar kinetic terms that are homogeneous and quadratic in
\(\mathbf b_+\) and \(\mathbf b_-\), we require rotational invariance,
forward--backward exchange symmetry, and recovery of the classical kinetic
energy. These conditions reduce the admissible terms to the one-parameter
family derived in Appendix~\ref{app:quadratic_kinetic_choice}:
\begin{equation}
\label{eq:general_symmetric_kinetic}
T_\kappa(\mathbf v,\mathbf u)
=
\frac{m}{2}
\left(
\mathbf v^{2}+\kappa\mathbf u^{2}
\right).
\end{equation}
Every value of \(\kappa\) in Eq.~\eqref{eq:general_symmetric_kinetic} has
the same classical limit \(\mathbf u\to0\). In the present work we choose
the member \(\kappa=-1\). By Eq.~\eqref{eq:b_pm_v}, its kinetic term has
the equivalent forms
\begin{equation}
\label{eq:bilinear_kinetic_identity}
T_{-1}
=
\frac{m}{2}
\left(
\mathbf v^{2}-\mathbf u^{2}
\right)
=
\frac{m}{2}\mathbf b_+\cdot\mathbf b_- .
\end{equation}
The corresponding Lagrangian is
\begin{equation}
\label{eq:lagrangian_symmetric}
L(\mathbf x,\mathbf b_+,\mathbf b_-,t)
=
\frac{m}{2}\mathbf b_+\cdot\mathbf b_-
-
V(\mathbf x,t).
\end{equation}
The kinetic term in Eq.~\eqref{eq:bilinear_kinetic_identity} is the
Guerra--Morato form~\cite{Guerra1983}. It is one admissible time-symmetric
choice within the family~\eqref{eq:general_symmetric_kinetic}, rather than
a consequence of time symmetry alone.

Appendix~\ref{app:quadratic_kinetic_choice} shows that, for
\(D=\hbar/(2m)\), recovery
of the standard Schr\"odinger equation selects \(\kappa=-1\) in the present
formulation. Although \(\kappa=+1\) is algebraically Yasue's kinetic
term~\cite{Yasue1981}, it enters a different variational problem: Yasue varies
the process with both mean derivatives as Lagrangian arguments, whereas here
only the forward drift is optimized after \(\mathbf u\) is expressed through
the density, with Eq.~\eqref{eq:forward_fp} as the state equation. In the
present formulation the \(\mathbf u^{2}\) contribution is control-independent
but changes the density costate equation, so the same kinetic expression need
not yield the same dynamics. Equation~\eqref{eq:lagrangian_symmetric} is
therefore a constitutive specification of the model, not a consequence of
time symmetry alone.

Substituting Eq.~\eqref{eq:backward_drift_relation} into
Eq.~\eqref{eq:lagrangian_symmetric}, the running cost becomes a functional of
the forward drift and the density,
\begin{equation}
\label{eq:law_dependent_running_cost}
L(\mathbf x,\mathbf b_+,P,t)
=
\frac{m}{2}\mathbf b_{+}^{2}
-
\frac{\hbar}{2}\mathbf b_+\cdot\nabla\log P
-
V(\mathbf x,t).
\end{equation}
The forward drift \(\mathbf b_+\) is the control field in the forward SDE,
while the density \(P\) is the law of the same controlled stochastic process.
Thus the forward--backward symmetric kinetic term induces a law-dependent
running cost.
Mathematically, this places the formulation in the class of mean-field-type, or
McKean--Vlasov-type, stochastic control problems: the cost depends on the law of
the controlled process. This terminology is rooted in Vlasov's self-consistent
kinetic theory and McKean's construction of Markov processes associated with
nonlinear parabolic equations~\cite{Vlasov1961,McKean1966}. Modern treatments
of dynamic programming and control for McKean--Vlasov systems can be found in~\cite{CarmonaDelarue2018,Pham2017}. Physically, however, \(P\) is not
the empirical density of many interacting particles. It is the probability
density of a single real diffusion, and its appearance follows from the
time-reversal relation between the forward and backward drifts.

\subsection{The density-space Bellman problem}

In the usual law-independent Markov-diffusion setting, the state of the control
problem is the particle position \(\mathbf x\), so the Bellman value is a
function \(J(\mathbf x,t)\) and satisfies a second-order HJB equation
\cite{Fleming2006}. Here the control problem is law-dependent, so the full
Bellman object is instead a value functional on the space of probability
densities. We denote it by
\begin{equation}
\label{eq:mean_field_value_functional}
\mathcal V(t,P).
\end{equation}
Here \(P\) denotes the whole density profile at time \(t\), so
\(\mathcal V\) is a functional on density space rather than a function of a
single position. For the running cost
\eqref{eq:law_dependent_running_cost},
\begin{equation}
\label{eq:mean_field_value_definition}
\mathcal V(t,P)
=
\inf_{\mathbf b_+}
\left\{
\int_t^{t_f}ds\int d^3x\,
P(\mathbf x,s)L(\mathbf x,\mathbf b_+,P,s)
+
\Phi[P(\cdot,t_f)]
\right\},
\end{equation}
where the infimum is over admissible forward drifts. In this control
problem, \(\mathbf b_+\) is the control field and \(P\) is the state
variable. For each admissible control, \(P\) evolves according to the forward
Fokker--Planck equation~\eqref{eq:forward_fp}, which therefore serves as the
state equation of the density-space control problem. The bilinear kinetic
term~\eqref{eq:bilinear_kinetic_identity} does not replace this density
evolution. Once the backward drift is expressed through the density of the
same diffusion, its specific role is to produce the law-dependent running
cost~\eqref{eq:law_dependent_running_cost}. The spatial integral
\(\int d^3x\,P L\) is the expectation of the running cost under the law \(P\),
so \(\mathcal V\) is the minimized expected value of the law-dependent action. The terminal functional
\(\Phi[P(\cdot,t_f)]\) enters only through the boundary condition
\(\mathcal V(t_f,P)=\Phi[P]\), not through the dynamic-programming equation
derived below. For the stationary problems considered in this paper, the
terminal cost plays no role and may be omitted.

Let us denote the local value
derivative by
\begin{equation}
\label{eq:local_value_derivative}
U(\mathbf x,t)
=
\frac{\delta\mathcal V}{\delta P}(t,P_t)(\mathbf x).
\end{equation}

The dynamic-programming equation on density space takes the form
\begin{align}
\label{eq:functional_hjb}
-\partial_t\mathcal V(t,P)
=
\inf_{\mathbf b_+}
\Bigg\{
&\int d^3x\,P L(\mathbf x,\mathbf b_+,P,t)
\nonumber\\
&+
\int d^3x\,U(\mathbf x,t)
\left[
-\nabla\cdot(\mathbf b_+P)
+
\frac{\hbar}{2m}\Delta P
\right]
\Bigg\}.
\end{align}
Appendix~\ref{app:functional_hjb_derivation} derives
Eq.~\eqref{eq:functional_hjb} from Bellman's principle on density space and
shows that, after integrating the Fokker--Planck terms by parts, the
minimization over \(\mathbf b_+\) becomes pointwise. The resulting optimality
condition is
\begin{equation}
\label{eq:optimal_forward_drift_U}
\mathbf b_+
=
-
\frac{1}{m}\nabla U
+
\frac{\hbar}{2m}\nabla\log P.
\end{equation}
It is useful to absorb the density-gradient term into the shifted local Bellman
field
\begin{equation}
\label{eq:shifted_bellman_field}
J
=
U
-
\frac{\hbar}{2}\log P.
\end{equation}
This shift is a change of local Bellman variable that absorbs the osmotic
density-gradient contribution into the gradient of \(J\). Then
Eq.~\eqref{eq:optimal_forward_drift_U} becomes
\begin{equation}
\label{eq:optimal_forward_drift}
\mathbf b_+
=
-
\frac{1}{m}\nabla J.
\end{equation}
Thus \(J\) is not the full mean-field value functional. It is the shifted local
Bellman field obtained from the functional derivative
\(U=\delta\mathcal V/\delta P\).

In terms of this shifted local Bellman field, the local law-dependent
HJB--Fokker--Planck system is
\begin{equation}
\label{eq:hjb_density_coupled_optimal}
-\partial_t J
=
-
V
-
\frac{1}{2m}|\nabla J|^2
+
\frac{\hbar}{2m}\Delta J
+
\frac{\hbar^2}{m}\frac{\Delta\sqrt P}{\sqrt P},
\end{equation}
together with
\begin{equation}
\label{eq:fp_optimal_J}
\partial_t P
=
\nabla\cdot\left(\frac{P}{m}\nabla J\right)
+
\frac{\hbar}{2m}\Delta P.
\end{equation}
Equation~\eqref{eq:hjb_density_coupled_optimal} is not a standard
law-independent HJB equation for a value function of \((\mathbf x,t)\) alone.
It is the shifted local HJB equation associated with the full density-space
value functional \(\mathcal V(t,P)\). The last term is the density contribution
generated by the law-dependent part of the coupled HJB--Fokker--Planck system.
In the reduction below, it combines with the remaining density terms to give
the standard quantum-potential term.

\subsection{Reduction to the Madelung and Schr\"odinger equations}

Combining Eqs.~\eqref{eq:probability_current_forward} and
\eqref{eq:optimal_forward_drift}, the current velocity is
\begin{equation}
\label{eq:current_velocity_from_J}
\mathbf v
=
-
\frac{1}{m}\nabla J
-
\frac{\hbar}{2m}\nabla\log P.
\end{equation}
This motivates the change of variables
\begin{equation}
\label{eq:value_function_definition}
S
=
-
J
-
\frac{\hbar}{2}\log P.
\end{equation}
Equivalently, since \(J=U-(\hbar/2)\log P\), one has \(S=-U\). Hence
\begin{equation}
\label{eq:current_velocity_from_S}
\mathbf v
=
\frac{1}{m}\nabla S,
\end{equation}
and the continuity equation becomes
\begin{equation}
\label{eq:continuity_ps}
\partial_t P
+
\nabla\cdot\left(P\frac{\nabla S}{m}\right)
=
0.
\end{equation}
As shown explicitly in Appendix~\ref{app:qhj_reduction}, substituting
Eq.~\eqref{eq:value_function_definition} into the shifted HJB equation
\eqref{eq:hjb_density_coupled_optimal}, and using the continuity equation
\eqref{eq:continuity_ps}, gives
\begin{equation}
\label{eq:quantum_hj_s}
\partial_t S
=
-
V
-
\frac{1}{2m}|\nabla S|^2
+
\frac{\hbar^2}{2m}\frac{\Delta\sqrt P}{\sqrt P}.
\end{equation}
Equations~\eqref{eq:continuity_ps} and~\eqref{eq:quantum_hj_s} are the
Madelung equations~\cite{Madelung1927}.

Introducing the wave function
\begin{equation}
\label{eq:hopf_cole_type}
\psi
=
\sqrt P\,e^{iS/\hbar},
\end{equation}
Eqs.~\eqref{eq:continuity_ps} and~\eqref{eq:quantum_hj_s} combine into the
Schr\"odinger equation
\begin{equation}
\label{eq:schrodinger_from_soc}
i\hbar\partial_t\psi
=
-
\frac{\hbar^2}{2m}\Delta\psi
+
V\psi.
\end{equation}
The density of the real diffusion is therefore identified with the Born density,
\begin{equation}
\label{eq:Born_rule}
P=|\psi|^2,
\end{equation}
and the phase by
\begin{equation}
\label{eq:phase_from_wavefunction}
S=\hbar\operatorname{Arg}\psi.
\end{equation}
Together with Eq.~\eqref{eq:hopf_cole_type}, these establish the Madelung correspondence between \((P,S)\) and \(\psi\) in both directions.
In the present stochastic optimal control formulation the wave function is not a primary object. The primary fields are the density \(P\) and the local value derivative (costate) \(U=\delta\mathcal V/\delta P\) defined in Eq.~\eqref{eq:local_value_derivative}. From these, the shifted field \(J\) of Eq.~\eqref{eq:shifted_bellman_field}, the Madelung phase \(S=-U\) of Eq.~\eqref{eq:value_function_definition}, and the wave function \(\psi\) of Eq.~\eqref{eq:hopf_cole_type} are reconstructed. The local HJB--Fokker--Planck system of Eqs.~\eqref{eq:hjb_density_coupled_optimal}--\eqref{eq:fp_optimal_J} for \((P,U)\) therefore admits a larger solution class than the Schr\"odinger equation: the standard single-valuedness condition along loops encircling nodes of \(\psi\)~\cite{Wallstrom1989,Wallstrom1994} selects the quantum subset of its solutions, and is imposed at the reconstruction step rather than derived from the local dynamics.
Away from the nodes of \(\psi\), Eq.~\eqref{eq:hopf_cole_type} together with
Eq.~\eqref{eq:b_pm_v} gives
\begin{equation}
\label{eq:forward_drift_wavefunction}
\mathbf b_+
=
\frac{\hbar}{m}\nabla\operatorname{Arg}\psi
+
\frac{\hbar}{m}\nabla\log|\psi|,
\end{equation}
the optimal feedback drift expressed in terms of the wave function. The Madelung correspondence then makes the two formulations operationally equivalent: in one direction, solving the HJB--Fokker--Planck system yields \((P,S)\) and hence the optimal drift directly, with \(\psi\) reconstructed via Eq.~\eqref{eq:hopf_cole_type}. In the other direction, solving the Schr\"odinger equation yields \(\psi\), from which \((P,S)\) follow via Eqs.~\eqref{eq:Born_rule} and~\eqref{eq:phase_from_wavefunction}, and the optimal drift via Eq.~\eqref{eq:forward_drift_wavefunction}. For the hydrogen eigenstates studied below, we exploit the second route: the closed-form wave functions \(\psi_{n\ell m}\) are inserted into Eq.~\eqref{eq:forward_drift_wavefunction} to obtain the drift used in the real-space simulations, bypassing the direct solution of the HJB--Fokker--Planck system.

For a time-independent potential \(V(\mathbf x)\), stationary solutions of
Eq.~\eqref{eq:schrodinger_from_soc} take the form
\[
\psi(\mathbf x,t)=\psi(\mathbf x)e^{-iEt/\hbar}.
\]
The time-dependent phase is spatially constant, so the spatial drift is
\begin{equation}
\label{eq:stationary_forward_drift}
\mathbf b_+(\mathbf x)
=
\frac{\hbar}{m}\nabla\operatorname{Arg}\psi(\mathbf x)
+
\frac{\hbar}{m}\nabla\log|\psi(\mathbf x)|,
\end{equation}
which is the form used in
Sections~\ref{sec:spherical-drift-central-potentials} and
\ref{sec:wavefunction-drift-hydrogen}.

From this point onward only the forward drift enters the analysis; we
therefore drop the subscript and write
\(\mathbf b\equiv\mathbf b_{+}\).


\section{Derivation of the Drift Velocity Field for Central Potentials in Spherical Coordinates}
\label{sec:spherical-drift-central-potentials}
	The study of Brownian motion on Riemannian manifolds is a rich field intersecting stochastic processes, differential geometry, and partial differential equations. This framework extends standard Brownian motion from Euclidean spaces to curved manifolds, capturing the manifold’s intrinsic geometry. In this work, Brownian motion will be analyzed within Euclidean spaces using curvilinear coordinate systems, where geometric factors arise from the coordinate choice rather than the space’s intrinsic curvature. Importantly, the mathematical formulations in both approaches are equivalent. Pioneering works by mathematicians such as It{\^o}~\cite{Ito1950,Ito1953,Ito1962}, J{\o}rgensen~\cite{Jorgensen1975}, Ikeda and Watanabe~\cite{Ikeda1981}, Elworthy~\cite{Elworthy1982}, Hsu~\cite{Hsu1988,Hsu2008}, and many others have laid the foundation for applying stochastic analysis to geometric contexts.

In this section, we build upon the findings of previous studies to derive the stochastic equation of motion for an electron in the field of a hydrogen nucleus.

Applications of stochastic mechanics to concrete systems have remained scarce. Nelson trajectories have been integrated numerically for several simple quantum systems --- the free particle, tunneling through a barrier, and model diatomic molecules --- by McClendon and Rabitz~\cite{McClendon1988}. For the hydrogen atom itself, the treatments we are aware of are that of Truman and Lewis~\cite{Truman1986}, who employed stochastic mechanics to analyze the ground state of the hydrogen atom, primarily focusing on the radial equation of motion and first hitting times, and the later analysis of the atomic elliptic state by Durran, Neate, and Truman~\cite{Durran2008DivineClockwork,Durran2008SpectralGap}, which exposes Keplerian motion in the semiclassical correspondence limit of the Nelson diffusion.
More recently, Carosso~\cite{Carosso2024} simulated Nelson trajectories for the \(1s\) and \(2s\) states and their superposition, displaying sample paths and the relaxation of the mean radial distance, in Cartesian coordinates with isotropic noise, as suffices for these \(\ell=0\) states.
None of these works treats eigenstates with nonzero angular momentum at the level of simulated trajectories, nor recovers the electron's energies or angular momentum from the trajectories themselves --- the regime studied here.

For the angular stochastic equations,
 Truman and Lewis~\cite{Truman1986} utilized the Stroock equation~\cite{Stroock1971} to model Brownian motion on the sphere $S^2$.
 However, they only provided the stochastic equation of motion on the sphere without delving deeper into these equations.
 Their approach involved constraining the process in $\mathbb{R}^3$ to remain on the sphere by incorporating projection and drift terms.
 In the current work, we express Brownian motion directly in terms of spherical coordinates, thereby capturing the motion intrinsically on $S^2$, based on the framework introduced by It{\^o}~\cite{Ito1962}. Although we do not adopt Stroock's approach~\cite{Stroock1971}, we acknowledge the existence of a third method to describe Brownian motion on a sphere, introduced by Price and Williams~\cite{Price1983}. For more details, refer to the work of van den Berg and Lewis~\cite{Berg1985}. Notably, both Stroock's and Price and Williams' approaches use extrinsic coordinates embedded in $\mathbb{R}^3$, unlike It{\^o}'s approach~\cite{Ito1962}, which explicitly incorporates the manifold's geometry through the metric tensor and Christoffel symbols.

Brownian motion on a Riemannian manifold $M$ is profoundly influenced by the manifold’s geometry. In stochastic process theory, the behavior of a stochastic process is characterized by its infinitesimal generator, which describes its limiting behavior over infinitesimally small time intervals. For Brownian motion on $M$, the infinitesimal generator is given by one half of the Laplace--Beltrami operator $\Delta_M$. Following the work of It{\^o}~\cite{Ito1962}, the generator of Brownian motion is expressed as:
\begin{equation}
\frac{1}{2} g^{i j} \nabla_i \nabla_j = \frac{1}{2} g^{i j} \frac{\partial^2}{\partial x^i \partial x^j} -  \frac{1}{2} g^{i j} \Gamma^k_{i j} \frac{\partial}{\partial x^k}.
\end{equation}

In It{\^o}’s original formulation, the diffusion coefficient is implicitly embedded within the metric tensor  $g^{ij}$. To make the diffusion coefficient explicit, we introduce a scalar diffusion coefficient  $\sigma$ and adjust the stochastic differential equation accordingly. We define $\sigma^i_k$  and $m^k$ by:
\begin{equation}
\label{eq:diffusion_coef_curved_sum}
\sigma^2 \sum_k \sigma^i_k \sigma^j_k = g^{i j},
\end{equation}
\begin{equation}
\label{eq:drift_velocity_curved_sum}
\sigma^2 m^k  = - \frac{1}{2} g^{i j} \Gamma^k_{i j},
\end{equation}
and solving a stochastic differential equation:
\begin{equation}
\label{eq:stochastic_equation_of_motion_ito}
dX^i  =  \sigma^2 m^i dt  + \sigma \sigma^i_k dW^k,
\end{equation}
we can describe the movement of the Brownian particle in curved coordinates.

If the metric tensor $g$ is diagonal, it can be shown that:
\begin{equation}
dX^i  =  \sigma^2 m^i dt  +  \sigma \sqrt{g^{i i}} dW^i.
\end{equation}

Introducing new notation for the diffusion coefficients and the drift velocity due to geometric curvature:
\begin{equation}
\label{eq:diffusion_coef_curved}
\sigma^i =  \sigma \sqrt{g^{i i}},
\end{equation}
\begin{equation}
\label{eq:drift_velocity_curved}
\mu^i =  \sigma^2 m^i,
\end{equation}
the stochastic equation of motion becomes:
\begin{equation}
dX^i  = \mu^i dt  +  \sigma^i dW^i.
\end{equation}

This form resembles the Cartesian case but with diffusion coefficients and geometric drift terms related to the metric tensor.

To account for the particle’s own forward drift $b^i$ in Euclidean space --- not resulting from curvature --- we add an additional drift term:
\begin{equation}
\label{eq:drift_with_particle_term}
dX^i  = \mu^i dt  + \sqrt{g^{i i}} b^i dt + \sigma^i dW^i.
\end{equation}

In Euclidean space with local spherical coordinates $(r, \theta, \phi)$, the non-zero components of the inverse metric tensor are:
\begin{equation}
\label{eq:metric_tensor}
g^{r r}=1, \qquad g^{\theta \theta} = \frac{1}{r^2}, \qquad g^{\phi \phi} = \frac{1}{r^2 \sin^2 \theta}.
\end{equation}

The diffusion coefficients in spherical coordinates, as given by Eq.~\eqref{eq:diffusion_coef_curved} become:
\begin{equation}
\sigma^r = \sigma \sqrt{g^{rr}} = \sigma, \qquad \sigma^\theta = \sigma \sqrt{g^{\theta\theta}} = \frac{\sigma}{r}, \qquad \sigma^\phi = \sigma \sqrt{g^{\phi\phi}} = \frac{\sigma}{r \sin \theta}.
\end{equation}

Since the metric tensor in spherical coordinates is diagonal, the non-zero Christoffel symbols involved are:
\begin{equation}
\label{eq:christoffel_symbols}
\begin{aligned}
\Gamma^r_{\theta \theta} = -r, \quad \Gamma^r_{\phi \phi} = -r \sin^2 \theta, \quad \Gamma^\theta_{\phi \phi} = - \sin \theta \cos \theta.
\end{aligned}
\end{equation}

Substituting the non-zero Christoffel symbols and the components of the inverse metric tensor into Eq.~\eqref{eq:drift_velocity_curved} for the drift velocity due to curvature, we obtain:
 \begin{equation}
\label{eq:geometric_drift_spherical}
\begin{aligned}
\mu^r= \sigma^2 \frac{1}{r}, \quad \mu^\theta = \sigma^2 \frac{\cot \theta}{2 r^2}, \quad \mu^\phi=0.
\end{aligned}
\end{equation}

Combining the diffusion and drift terms arising from the geometry of the spherical coordinate system, the stochastic equations governing Brownian motion in spherical coordinates are:
\begin{equation}
\label{eq:stochastic_equation_of_motion_spherical}
\begin{aligned}
dr &= \left( \frac{\sigma^2}{r} + b_r \right) dt + \sigma dW^r, \\
d\theta &= \left( \frac{\sigma^2 \cot \theta}{2 r^2} + \frac{1}{r} b_\theta \right) dt + \frac{\sigma}{r} dW^\theta, \\
d\phi &= \frac{1}{r \sin \theta} b_\phi\, dt + \frac{\sigma}{r \sin\theta} dW^\phi,
\end{aligned}
\end{equation}
where $b_r$, $b_\theta$, and $b_\phi$ are the spherical components of the forward drift $\mathbf b$.


\section{Wave Function-Based Calculation of Drift Fields for Electrons in Central Potentials}
\label{sec:wavefunction-drift-hydrogen}
It is well known~\cite{Griffiths2018} that the solution to the stationary Schrödinger equation in spherical coordinates for the electron in a central potential is given by:
\begin{equation}
\label{eq:psi_R_Y}
\psi_{n\ell m}(r, \theta, \phi) = R_{n\ell}(r) Y_{\ell m}(\theta, \phi),
\end{equation}
where $R_{n\ell}(r)$ is the radial wave function and $Y_{\ell m}(\theta, \phi)$ represents the spherical harmonics, which constitute the angular part of the wave function.

The spherical harmonics, $Y_{\ell m}(\theta, \phi)$, depend on the angular momentum quantum numbers $\ell$ and $m$ and can be expressed as:
\begin{equation}
\label{eq:angular_wave_function_theta_phi}
Y_{\ell m}(\theta, \phi) = \Theta_{\ell m}(\theta) \Phi_m(\phi),
\end{equation}
where $\Theta_{\ell m}(\theta)$ are the normalized associated Legendre polynomials $P_{\ell}^m(\cos\theta)$, and $\Phi_m(\phi)$ is the azimuthal function.

The explicit form of the hydrogenic wave function used in the present work is given in Appendix~\ref{app:wave_function_hydrogen}.

In this paper, we will typically use the notation $(n,\ell,m)$ to represent the quantum state of the electron. For example, the state with $n=2,\ell=1$, and $m=0$ will be written as $(2,1,0)$. From this section onward, the symbol $m$ refers to the magnetic quantum number; the electron mass is denoted $m_e$.

For a stationary hydrogen eigenstate, the forward drift $\mathbf b$ is given by Eq.~\eqref{eq:stationary_forward_drift}. Using Eqs.~\eqref{eq:psi_R_Y} and~\eqref{eq:angular_wave_function_theta_phi}, together with the hydrogenic wave function given in Appendix~\ref{app:wave_function_hydrogen}, we note that the radial and polar factors \(R_{n\ell}(r)\) and \(\Theta_{\ell m}(\theta)\) are real, so the phase of \(\psi\) varies with position only through the azimuthal factor \(\Phi_m(\phi)=e^{im\phi}\), whose argument is \(\operatorname{Arg}\Phi_m=m\phi\). Hence,
\begin{equation}
\mathbf b=\frac{\hbar}{m_e} \nabla \log \left| R_{n\ell}(r)\Theta_{\ell m}(\theta)\right|+\frac{\hbar}{m_e}\nabla \operatorname{Arg}\Phi_m(\phi).
\end{equation}

Applying the gradient operator in spherical coordinates, we derive:
\begin{equation}
\mathbf b= \frac{\hbar}{m_e}  \frac{d}{d r} \log \left| R_{n\ell}(r) \right| \hat r + \frac{\hbar}{m_e}  \frac{1}{r}  \frac{d}{d \theta}\log \left| \Theta_{\ell m}(\theta) \right| \hat \theta + \frac{\hbar}{m_e} \frac{1}{r \sin{\theta}} \frac{d}{d \phi} \operatorname{Arg}\Phi_m(\phi) \hat \phi.
\end{equation}

The final expression for the forward drift field $\mathbf{b} = (b_r, b_\theta, b_\phi)$ is:
\begin{equation}
\label{eq:drift_velocity_spherical}
\begin{aligned}
b_r &= \frac{\hbar}{m_e} \frac{d}{dr}\log \left| R_{n\ell}(r) \right|, \\
b_\theta &=  \frac{\hbar}{m_e}  \frac{1}{r}  \frac{d}{d \theta}\log \left| \Theta_{\ell m}(\theta) \right|, \\
b_\phi &=  \frac{\hbar}{m_e} \frac{1}{r \sin{\theta}} \frac{d}{d \phi}\operatorname{Arg}\Phi_m(\phi).
\end{aligned}
\end{equation}

For hydrogen eigenstates the amplitude
\(R_{n\ell}(r)\,\Theta_{\ell m}(\theta)\) is independent of the azimuthal angle
\(\phi\), so the azimuthal component of the osmotic
velocity~\eqref{eq:osmotic_velocity} vanishes, \(u_\phi=0\), and by
Eq.~\eqref{eq:b_pm_v} the azimuthal
drift coincides with the current velocity,
\(b_\phi=v_\phi=\hbar m/(m_e r\sin\theta)\). For \(m\neq 0\) this is a
non-vanishing purely-phase-driven azimuthal current that persists in the
stationary state.

\section{Calculating Average Radial Kinetic Energy Using a Stochastic Approach}
\label{sec:avg_radial_kinetic_energy}
In quantum mechanics, the kinetic energy of an electron is determined by applying the kinetic energy operator and computing its expectation value through the corresponding integral (see Eq.~\eqref{eq:expectation_value_of_kinetic_energy}). For a detailed derivation of the kinetic energy using the operator approach, refer to Appendix~\ref{sec:avg_radial_kinetic_energy_operator}.

In the present real-valued stochastic optimal control framework, we define the kinetic energy of the electron not as a Hermitian operator but as the classical kinetic energy of a particle:
\begin{equation}
\label{eq:radial_kinetic_energy}
T_r = \frac{1}{2} m_e b_r^2.
\end{equation}

For the stationary hydrogen eigenstates considered here, the radial and polar components of \(\mathbf b\) are purely osmotic and its azimuthal component is purely phase-driven (Section~\ref{sec:wavefunction-drift-hydrogen}), so that \(\mathbf u\cdot\mathbf v=0\) and \(\tfrac{1}{2}m_e\mathbf b^{2}=\tfrac{1}{2}m_e\left(\mathbf u^{2}+\mathbf v^{2}\right)\). The classical kinetic energy of the drift therefore coincides with the time-symmetric kinetic energy of the stochastic variational formulations~\cite{Yasue1981}.

In this section, we show that the operator approach and the stochastic approach yield identical results for the average radial kinetic energy.
In the subsequent sections, we employ numerical simulations to compute the average kinetic energy and show that it matches the values calculated using Eqs.~\eqref{eq:expectation_value_of_kinetic_energy} and~\eqref{eq:avg_value_of_kinetic_energy}.

The average radial kinetic energy is then calculated by taking the expectation value of $T_r$ with respect to the probability density of the electron’s position. This is expressed as:
\begin{equation}
\label{eq:avg_value_of_kinetic_energy}
\langle T_r \rangle = \int_0^{\infty} \int_0^{\pi} \int_0^{2\pi} \left| R_{n\ell}(r) Y_{\ell m}(\theta, \phi) \right|^2 T_r \, r^2 \sin \theta \, d\phi \, d\theta \, dr.
\end{equation}

Separating the angular part, we obtain:
\begin{equation}
\langle T_r \rangle = \frac{1}{2} m_e \int_0^\infty r^2 R^2_{n\ell}(r) b_r^2 \, dr \int_0^{2\pi} \int_0^\pi \left|Y_{\ell m}(\theta, \phi)\right|^2 \sin \theta \, d\theta \, d\phi.
\end{equation}

Using the normalization of spherical harmonics, the angular integral simplifies to unity:
\begin{equation}
\langle T_r \rangle = \frac{\hbar^2}{2 m_e}  \int_0^\infty r^2 R^2_{n\ell}(r) \left(\frac{1}{R_{n\ell}(r)} \frac{d R_{n\ell}(r)}{dr} \right)^2 \, dr.
\end{equation}

Using the identity
\(\bigl(\tfrac{d(rR_{n\ell})}{dr}\bigr)^{2}
=r^{2}\bigl(\tfrac{dR_{n\ell}}{dr}\bigr)^{2}+\tfrac{d}{dr}\bigl(rR_{n\ell}^{2}\bigr)\)
together with \(\bigl[rR_{n\ell}^{2}\bigr]_{0}^{\infty}=0\), which holds for
bound states, the expression for the average radial kinetic energy reduces to:
\begin{equation}
\label{eq:radial_KE_stochastic}
\langle T_r \rangle = \frac{\hbar^2}{2 m_e}  \int_0^\infty \left(\frac{d ( r R_{n\ell}(r))}{dr} \right)^2 \, dr.
\end{equation}

We will now demonstrate that the average radial kinetic energy calculated using the stochastic approach in Eq.~\eqref{eq:radial_KE_stochastic} is equal to that obtained via the operator approach in Eq.~\eqref{eq:radial_KE}.

Let us introduce the notation:
\begin{equation}
u(r)= r R, \qquad u^{\prime}(r) = R + r R^{\prime}, \qquad u^{\prime\prime}(r)= 2R^{\prime} + r R^{\prime\prime}.
\end{equation}

Next, we perform integration by parts on the following integral:
\begin{equation}
\int_{0}^{\infty} [u^{\prime}(r)]^2 dr = \int_{0}^{\infty}  u^{\prime}(r) u^{\prime}(r) dr = u^{\prime}(r) u(r) \Big |_{0}^{\infty}  - \int_{0}^{\infty} u(r) u^{\prime \prime}(r) dr.
\end{equation}

Assuming that $R(r)$ vanishes sufficiently rapidly as $r \to 0$ and $r \to \infty$, and that its derivatives remain finite, we obtain:
\begin{equation}
u^{\prime}(r) u(r) \Big|_{0}^{\infty} = 0.
\end{equation}

Consequently, the integral simplifies to:
\begin{equation}
\label{eq:radial_integral}
\int_{0}^{\infty} [u^{\prime}(r)]^2 \, dr = - \int_{0}^{\infty} u(r) u^{\prime\prime}(r) \, dr.
\end{equation}

The left-hand side of Eq.~\eqref{eq:radial_integral} represents the radial kinetic energy obtained through the stochastic approach:
\begin{equation}
\int [u^{\prime}(r)]^2 dr = \int \left(\frac{d (r R)}{dr}\right)^2 dr.
\end{equation}

Conversely, the right-hand side of Eq.~\eqref{eq:radial_integral} corresponds to the average kinetic energy obtained via the operator approach:
\begin{equation}
\int u(r) u^{\prime\prime}(r) dr  = \int r R (2 R^{\prime} + r R^{\prime\prime}) dr = \int R (2 r R^{\prime} + r^2 R^{\prime\prime}) dr = \int R \frac{d}{dr}\left( r^2  \frac{d R}{dr} \right) dr.
\end{equation}

Finally, we have demonstrated that the average kinetic energy calculated using the stochastic approach in Eq.~\eqref{eq:radial_KE_stochastic} is equal to that obtained via the operator approach in Eq.~\eqref{eq:radial_KE}:
\begin{equation}
\int \left(\frac{d (r R)}{dr}\right)^2 dr  = - \int R \frac{d}{dr}\left( r^2  \frac{d R}{dr} \right) dr.
\end{equation}

\section{Calculating Average Angular Kinetic Energy Using a Stochastic Approach}
\label{sec:avg_angular_kinetic_energy}
In quantum mechanics, the angular kinetic energy of an electron is determined by applying the angular kinetic energy operator and computing its expectation value through the corresponding integral (see Eq.~\eqref{eq:expectation_value_of_angular_kinetic_energy}). For a detailed derivation of the angular kinetic energy using the operator approach, refer to Appendix~\ref{sec:avg_angular_kinetic_energy_operator}.

As in Section~\ref{sec:avg_radial_kinetic_energy}, we define the angular kinetic energy of the electron not in terms of a Hermitian operator, but as the classical kinetic energy associated with the particle’s angular motion:
\begin{equation}
T_{\text{angular}} = T_\theta + T_\phi,
\end{equation}
where $T_\theta$ and $T_\phi$ are respectively the polar and azimuthal kinetic energies, given by:
\begin{equation}
\label{eq:polar_and_azimuthal_kinetic_energies}
T_\theta = \frac{1}{2} m_e b_\theta^2, \quad T_\phi = \frac{1}{2} m_e b_\phi^2.
\end{equation}

The average angular kinetic energy is then calculated by taking the expectation value of $T_{\text{angular}}$ with respect to the probability density of the electron’s position. This is expressed as:
\begin{equation}
\langle T_{\text{angular}} \rangle =  \frac{1}{2} m_e \int_0^{\infty} \int_0^{\pi} \int_0^{2\pi} \left| R_{n\ell}(r)  Y_{\ell m}(\theta, \phi) \right|^2 \left(b_\theta^2  + b_\phi^2\right) \, r^2 \sin \theta \, d\phi \, d\theta \, dr.
\end{equation}

We first compute the polar kinetic energy, which is associated with the motion in the $\theta$ direction:
\begin{equation}
\langle T_\theta \rangle = \frac{1}{2} m_e  \int_0^\infty \left[ R_{n\ell}(r) \right]^2 r^2 \, dr \, 2 \pi \int_0^\pi \sin(\theta)  \left|\Theta_{\ell m}(\theta)\right|^2 b_\theta^2 \, d \theta.
\end{equation}

Using Eq.~\eqref{eq:drift_velocity_spherical}, we obtain:
\begin{equation}
\label{eq:avg_polar_kinetic_energy}
\langle T_\theta \rangle = \frac{\hbar^2}{2 m_e} \left\langle \frac{1}{r^2} \right\rangle 2 \pi \int_0^\pi \sin(\theta)  \, \left|\Theta_{\ell m}(\theta) \right|^2 \left( \frac{1}{\Theta_{\ell m}(\theta)}  \frac{d \Theta_{\ell m}(\theta)}{d \theta}  \right)
^2 \, d \theta.
\end{equation}

Using the normalization of \(P_{\ell}^{m}\), the relation \(\int|\nabla_{\Omega}Y_{\ell m}|^{2}\,d\Omega=\ell(\ell+1)\) for the angular gradient on the unit sphere, and Eq.~\eqref{eq:integral_squrate_legandre_poly} below, one obtains:
\begin{equation}
\label{eq:integral_squrate_derivative_legandre_poly}
\int_0^\pi \sin(\theta)  \, \left( \frac{d P_{\ell}^m(\cos\theta)}{d \theta}  \right) ^2 \, d \theta =\frac{(\ell+|m|)!}{(\ell-|m|)!}\, \frac{2 \ell^2 - 2 \ell (|m|-1) - |m|}{2 \ell+1}.
\end{equation}

Using Eqs.~\eqref{eq:normalized_legendre_poly}, \eqref{eq:integral_squrate_derivative_legandre_poly}, and~\eqref{eq:avg_polar_kinetic_energy}, we obtain a derivation of the electron’s average polar kinetic energy in a central potential using a stochastic approach:
\begin{equation}
\label{eq:avg_polar_kinetic_energy_final}
\langle T_\theta \rangle = \frac{\hbar^2}{2 m_e} \left\langle \frac{1}{r^2} \right\rangle
 \frac{2 \ell^2 - 2 \ell (|m|-1) - |m|}{2}.
\end{equation}

Next, we compute the azimuthal kinetic energy, which is associated with the motion in the $\phi$ direction:
\begin{equation}
\langle T_\phi \rangle = \frac{1}{2} m_e  \int_0^\infty  \left[ R_{n\ell}(r) \right]^2 r^2 \, dr \, \int_0^\pi \sin(\theta) \left|\Theta_{\ell m}(\theta)\right|^2 \, d \theta  \int_0^{2 \pi} \left|\Phi_m(\phi)\right|^2 b_\phi^2 \, d \phi.
\end{equation}

Using Eq.~\eqref{eq:drift_velocity_spherical}, we obtain:
\begin{equation}
\label{eq:avg_azimuthal_kinetic_energy}
\langle T_\phi \rangle = \frac{\hbar^2}{2 m_e} \left\langle \frac{1}{r^2} \right\rangle \, \int_0^\pi \frac{1}{\sin(\theta)} \left|\Theta_{\ell m}(\theta)\right|^2 \, d \theta  \int_0^{2 \pi} \left|\Phi_m(\phi)\right|^2 \left( \frac{d}{d \phi}\operatorname{Arg}\Phi_m(\phi) \right)^2 \, d \phi.
\end{equation}

For \(m\neq 0\), the associated Legendre functions satisfy the orthogonality relation~\cite{arfken2013}:
\begin{equation}
\label{eq:integral_squrate_legandre_poly}
\int_0^\pi \frac{1}{\sin(\theta)} \left[P_{\ell}^m(\cos\theta)\right]^2 \, d \theta = \frac{(\ell+|m|)!}{(\ell-|m|)!} \, \frac{1}{|m|}.
\end{equation}

Using Eqs.~\eqref{eq:normalized_legendre_poly}, \eqref{eq:integral_squrate_legandre_poly}, and~\eqref{eq:avg_azimuthal_kinetic_energy}, we calculate the azimuthal kinetic energy as:
\begin{equation}
\langle T_\phi \rangle = \frac{\hbar^2}{2 m_e} \left\langle \frac{1}{r^2} \right\rangle \, \frac{2 \ell + 1}{4 \pi} \frac{1}{|m|}  \int_0^{2 \pi} \left( \frac{d}{d \phi}\operatorname{Arg}\Phi_m(\phi) \right)^2 \, d \phi.
\end{equation}

For the azimuthal factor defined in Eq.~\eqref{eq:phi_m_complex}, one has $\frac{d}{d \phi}\operatorname{Arg}\Phi_m(\phi) = m$. Consequently,
\begin{equation}
\int_0^{2 \pi} \left( \frac{d}{d \phi}\operatorname{Arg}\Phi_m(\phi) \right)^2 \, d \phi = 2\pi m^2,
\end{equation}
and the final result for the azimuthal kinetic energy is:
\begin{equation}
\label{eq:avg_azimuthal_kinetic_energy_final}
\langle T_\phi \rangle = \frac{\hbar^2}{2 m_e} \left\langle \frac{1}{r^2} \right\rangle \, \frac{2 \ell + 1}{2} \, |m|.
\end{equation}
For \(m=0\), one has \(b_\phi=0\), so \(\langle T_\phi\rangle=0\).

Using the equation for the polar kinetic energy~\eqref{eq:avg_polar_kinetic_energy_final} and the equation for the azimuthal kinetic energy~\eqref{eq:avg_azimuthal_kinetic_energy_final}, we calculate the total angular kinetic energy as:
\begin{equation}
\label{eq:avg_angular_kinetic_energy_final}
\langle T_{\text{angular}} \rangle =\langle T_\theta \rangle  +  \langle T_\phi \rangle = \frac{\ell (\ell + 1) \hbar^2}{2 m_e} \left\langle \frac{1}{r^2} \right\rangle.
\end{equation}

While the average polar and azimuthal kinetic energies depend on the magnetic quantum number $m$, this dependence cancels out when summing both energies to determine the total angular kinetic energy. Consequently, the result for the angular kinetic energy obtained using the stochastic approach matches the result obtained via the operator approach in Eq.~\eqref{eq:avg_angular_energy_operator}.


\section{Stochastic Simulation of the Hydrogen Atom}
\label{sec:hydrogen-simulation}
In the present study, we develop a computer simulation of the real-valued
controlled stochastic process derived in
Section~\ref{sec:soc-fp-foundations} for the hydrogen atom. The numerical
simulation described below does not solve the
density-space Bellman equation directly. Instead, for stationary hydrogen
eigenstates we exploit the equivalence between the two formulations established
in Section~\ref{sec:soc-fp-foundations}: rather than solving the local
HJB--Fokker--Planck system of Eqs.~\eqref{eq:hjb_density_coupled_optimal}--\eqref{eq:fp_optimal_J}
for \((P,U)\), the closed-form wave functions \(\psi_{n\ell m}\) of the
Schr\"odinger equation are inserted into
Eq.~\eqref{eq:forward_drift_wavefunction} to obtain the feedback drift. The
simulated trajectories are therefore realizations of the real-valued controlled
stochastic process of Section~\ref{sec:soc-fp-foundations}, with forward drift
given by the optimal feedback, rather than realizations of an independently
postulated stochastic process. With sufficient sampling, the empirical
spherical-coordinate distributions of the trajectory converge to the
corresponding Born marginals predicted by the wave function.

The simulation algorithm is straightforward. We begin by initializing the position of a single electron near the hydrogen nucleus. At each time step, the simulation updates the electron’s position using the stochastic equation of motion in spherical coordinates~\eqref{eq:stochastic_equation_of_motion_spherical} and the corresponding forward drift computed from Eq.~\eqref{eq:drift_velocity_spherical}. Alternatively, we merge both equations into a single combined stochastic equation of motion for the electron in the hydrogen atom:
\begin{equation}
\begin{aligned}
\label{eq:final_stochastic_equation_of_motion}
dr &=  \frac{\hbar}{m_e}  \left( \frac{1}{r} + \frac{d}{dr}\log \left| R_{n\ell}(r) \right| \right) dt + \sigma dW^r, \\
d\theta &= \frac{1}{r^2} \frac{\hbar}{m_e} \left( \frac{\cot \theta}{2} + \frac{d}{d \theta}\log \left| \Theta_{\ell m}(\theta) \right| \right) dt + \frac{\sigma}{r} dW^\theta, \\
d\phi &= \frac{\hbar\,m}{m_e r^2 \sin^2 \theta} \, dt + \frac{\sigma}{r \sin\theta} dW^\phi.
\end{aligned}
\end{equation}

The stochastic equations~\eqref{eq:final_stochastic_equation_of_motion}
were integrated using the Euler--Maruyama scheme with a fixed time step
\(\Delta t=10^{-20}\,{\rm s}\)
(\(\Delta t=5\times10^{-21}\,{\rm s}\) for the \((2,1,m)\) runs shown in
Figures~\ref{fig:distributions_2p1}, \ref{fig:azimuthal_current_2p1},
and~\ref{fig:lz_running_estimator}, and for the \((2,1,0)\) energy run in
the left panel of Figure~\ref{fig:energies_2p0_2s0}), more than three orders
of magnitude below the
natural hydrogen time scale
\(a_0/v_B = a_0/(\alpha c)\approx 2.4\times 10^{-17}\,{\rm s}\). The
corresponding typical drift displacement per step,
\(v_B\,\Delta t\sim 4\times 10^{-4}\,a_0\), and Brownian displacement,
\(\sqrt{2D\,\Delta t}\sim 2\times 10^{-2}\,a_0\), are both well below the
orbital scale. Unless stated otherwise, the distribution plots were obtained
from one long trajectory of a single particle; the stability analysis in
Section~\ref{sec:simulation_convergence_2p0} uses multiple trajectories, as
specified there. The random number generator was initialized with a fixed seed
in order to make the numerical results reproducible.

The coordinate processes use reflecting boundary conditions at the origin
(\(r\leftarrow|r|\)) and at the poles (\(\theta=0,\pi\)), and a periodic
condition in the azimuth \(\phi\). The drift is singular at the origin, the
poles, and the wave-function nodes; its regularization, and the resulting
convergence of the energy averages, are analyzed in
Section~\ref{sec:simulation_convergence_2p0}.

We visualize the electron’s trajectory in three dimensions and collect statistical data on the radial and angular distributions of its position. After a sufficient simulation time, these distributions converge to the probability distributions~\eqref{eq:probability_distributions} predicted by the radial and angular wave functions~\eqref{eq:radial_wave_function} and~\eqref{eq:angular_wave_function} and Born rule, respectively. Additionally, we calculate the average energies of the electron and compare our numerical results with the analytical solutions detailed in Sections~\ref{sec:avg_radial_kinetic_energy} and~\ref{sec:avg_angular_kinetic_energy}. For the states with \(m\neq0\) we also track the phase-driven azimuthal circulation and the running time-average of the angular momentum accumulated along the trajectory, which converges to the quantized value \(L_z=m\hbar\) (Section~\ref{subsec:azimuthal-current}).

The source code for the simulation is available on GitHub (see Reference~\cite{github_Yordanov2024}).
The repository README provides the parameter sets and commands needed to run
the simulations and reproduce the figures.

\section{3D Visualization of Electron Trajectories in the Hydrogen Atom}
\label{sec:hydrogen-trajectories-3d}
\begin{figure}[h!]
 \vspace{0pt}
    \centering
    \begin{minipage}[b]{0.49\textwidth}
       \centering
       \includegraphics[width=\textwidth]{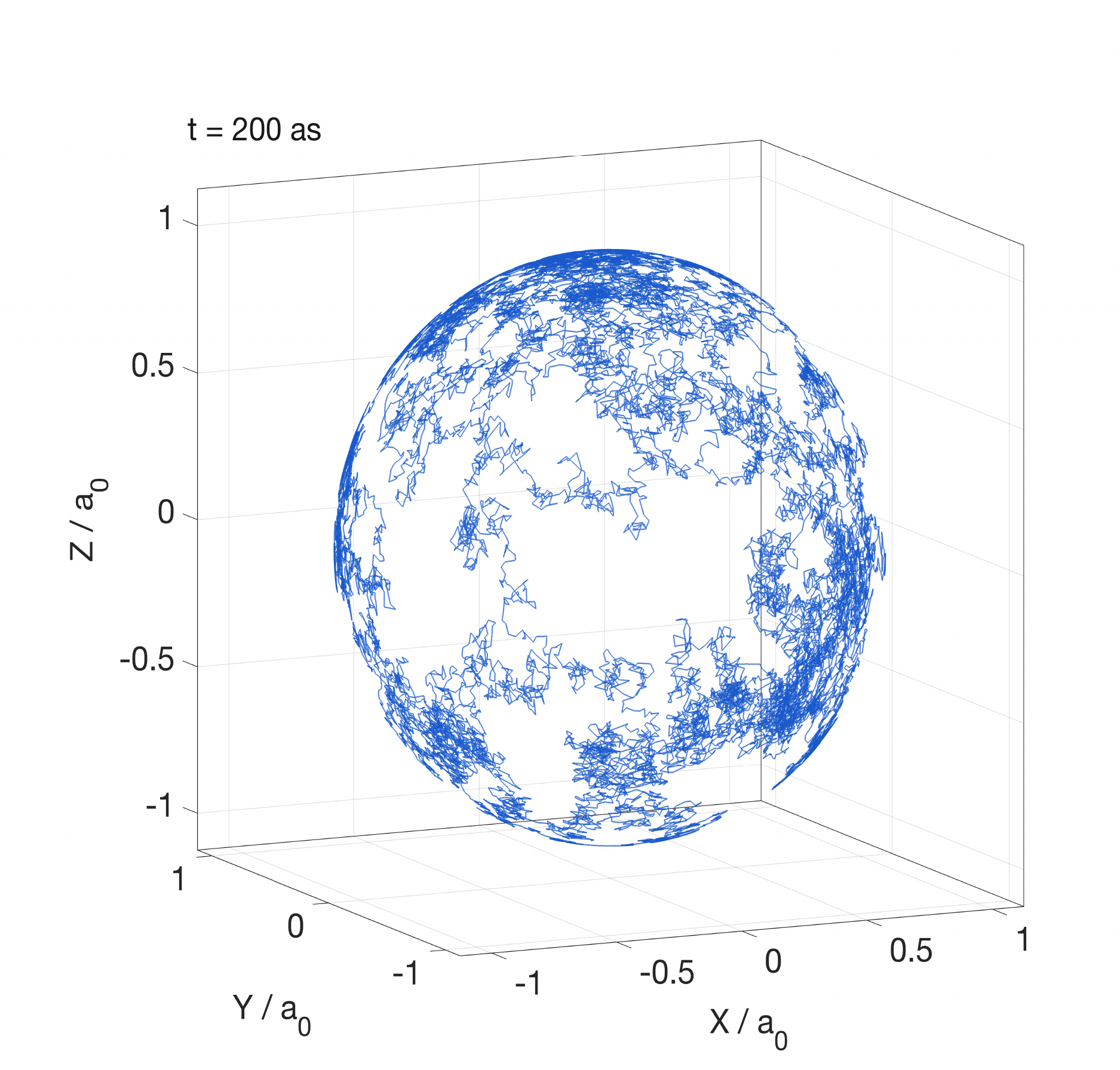}
    \end{minipage}
    \begin{minipage}[b]{0.49\textwidth}
       \centering
       \includegraphics[width=\textwidth]{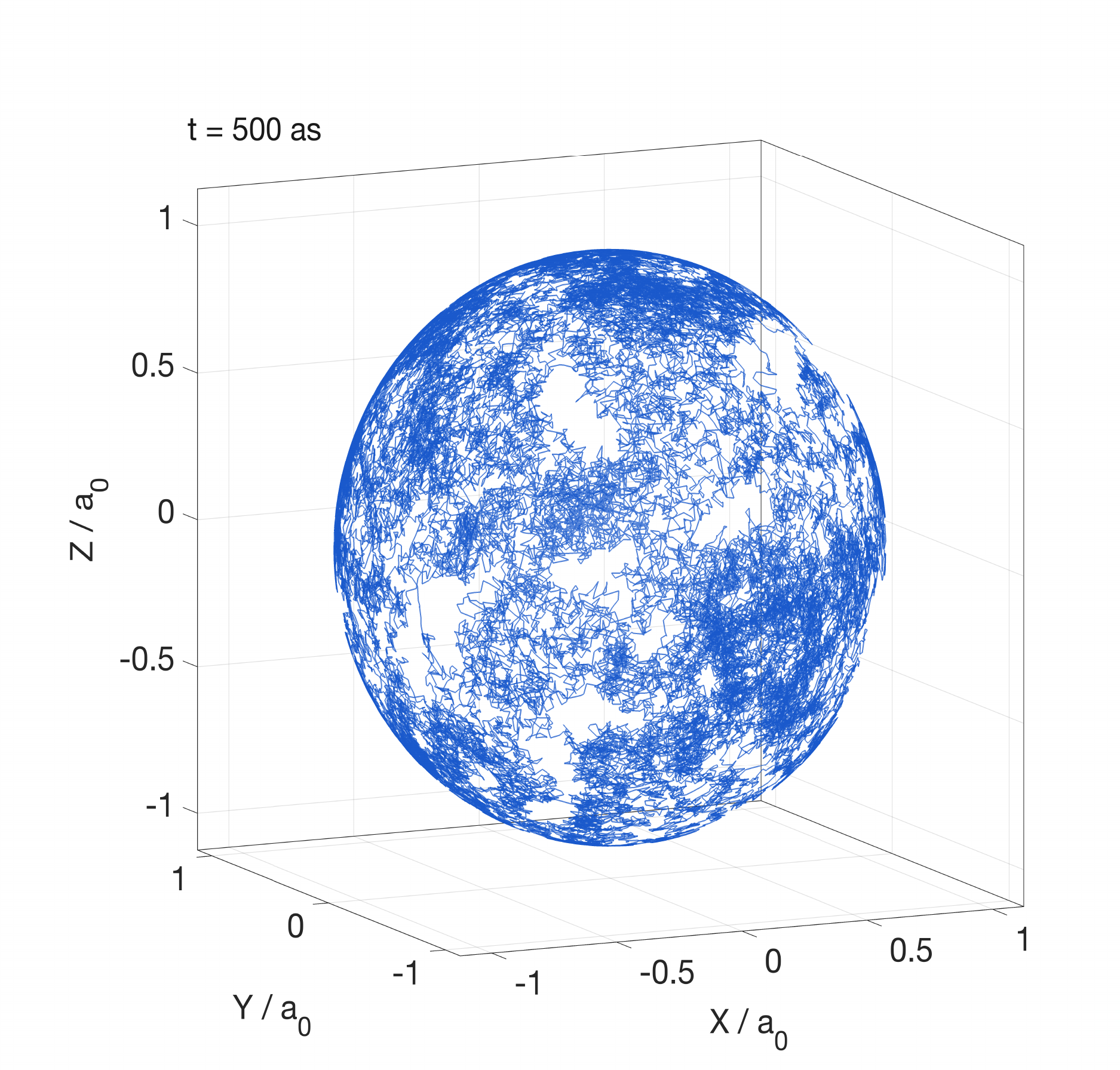}
    \end{minipage}
    \vspace{-8pt}
    \caption{Stochastic trajectory of the electron constrained to a spherical surface. For demonstration purposes, the radial diffusion coefficient is set to zero ($\sigma_r = 0$), resulting in motion confined to a sphere with radius $a_0$ (Bohr radius). The electron experiences the forward drift field $\mathbf{b} = \frac{\hbar}{m_e} \,\nabla \log |\psi_{1,0,0}|$. Initially, the electron is positioned on the sphere at spherical coordinates $(a_0, \frac{\pi}{2}, 0)$.
}
\label{fig:brownian_motion_sphere}
\end{figure}

To provide intuition for the reader regarding the stochastic trajectories of the electron within the hydrogen atom, we present an artificial trajectory of the electron in Figure~\ref{fig:brownian_motion_sphere}. In this simulation, we set the stochastic term in the radial stochastic equation of motion~\eqref{eq:stochastic_equation_of_motion_spherical} to zero and simulate the $(1,0,0)$ state by initializing the electron’s radial coordinate to the Bohr radius, $a_0 = 5.29177 \times 10^{-11} \, \text{m}$.

\begin{figure}[h!]
    \centering
    \begin{minipage}[b]{0.49\textwidth}
       \centering
       \includegraphics[width=\textwidth]{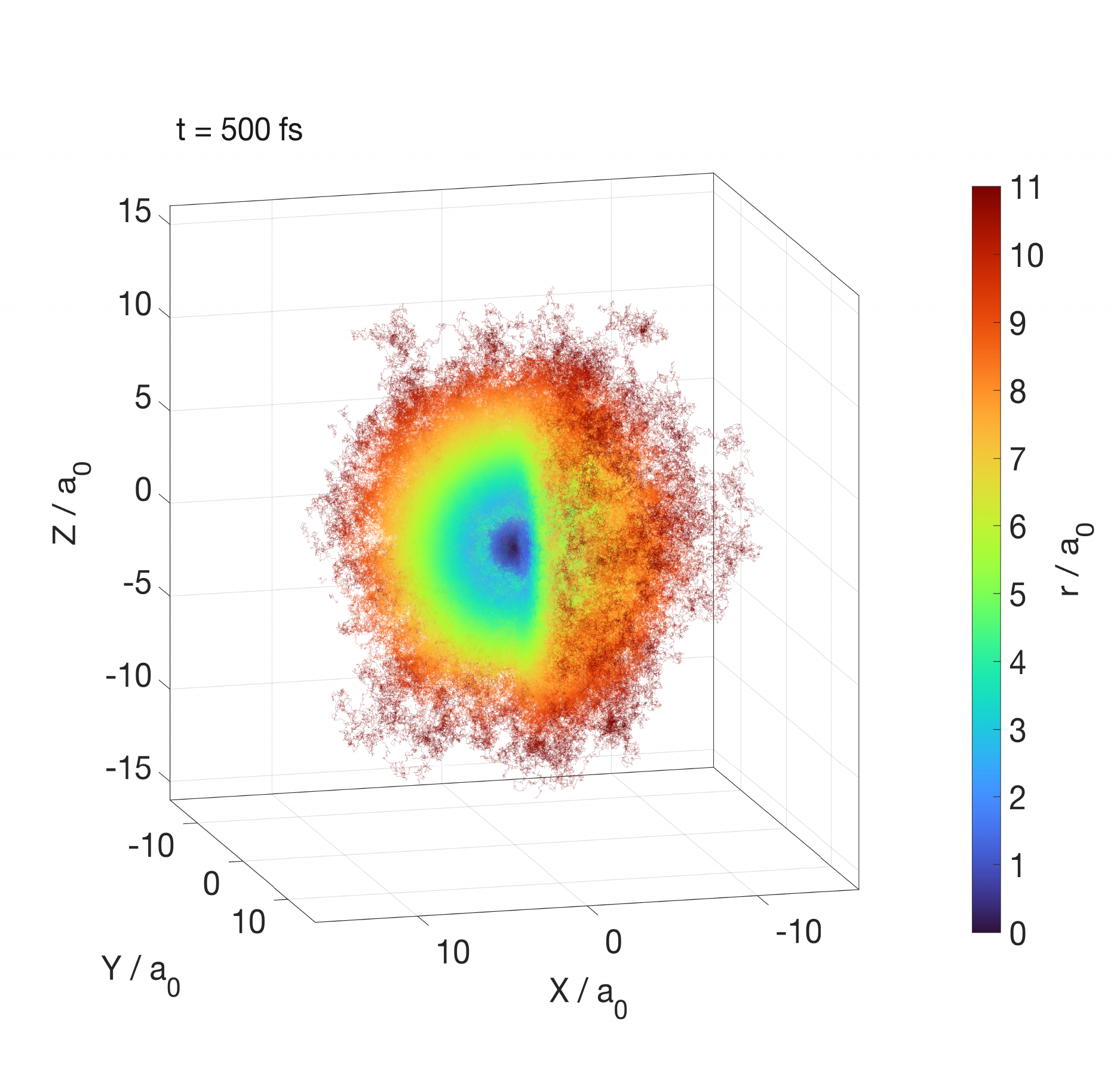}
    \end{minipage}\hfill
    \begin{minipage}[b]{0.49\textwidth}
       \centering
       \includegraphics[width=\textwidth]{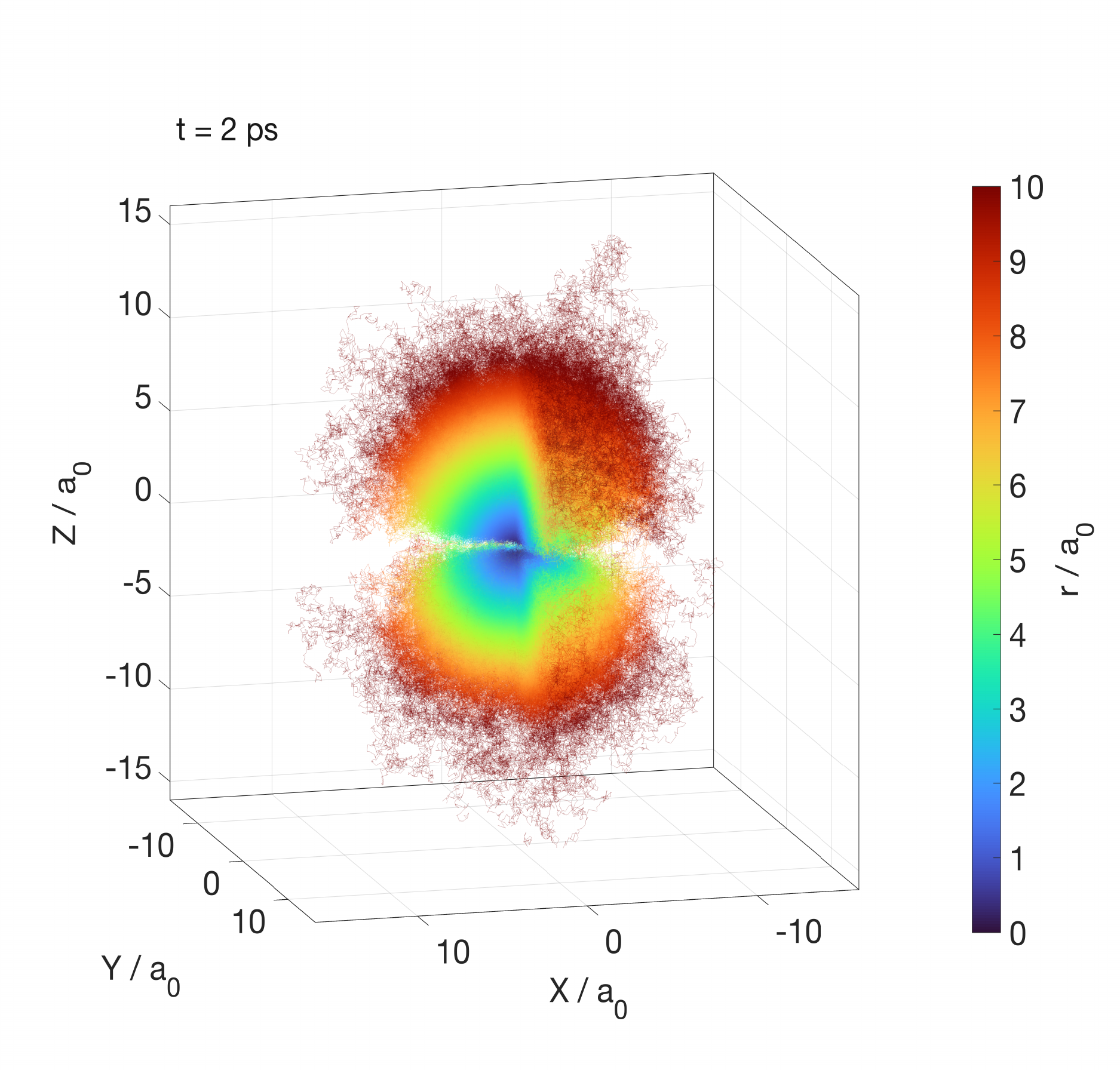}
    \end{minipage}
    \vspace{-7pt}
    \caption{Stochastic trajectory of the electron in three-dimensional space. The trajectory is colored by the radial distance $r/a_0$. In both panels the quarter-space $x>0$, $y>0$ is cut away to expose the interior of the cloud.
    \textbf{Left:} Forward drift field of the electron is: $\mathbf{b} = \frac{\hbar}{m_e} \, \nabla \log|\psi_{2,0,0}|$. The radial node at $r=2a_0$ separates the compact inner core from the diffuse outer shell. \textbf{Right:} Forward drift field of the electron is: $\mathbf b = \frac{\hbar}{m_e} \, \nabla \log |\psi_{2,1,0}|$. The nodal plane $\theta=\pi/2$ between the two lobes is clearly visible.}
\label{fig:3d_electron_in_h_atom}
\end{figure}

This simplification artificially suppresses the radial motion, enabling the visualization of the electron’s stochastic trajectory on a fixed spherical surface. Similar trajectories of particles constrained to a sphere have been reported in other studies of Brownian motion on spherical geometries~\cite{besser2023, angst2015}. However, it is important to note that in reality, the stochastic dynamics encompasses not only movement along a fixed radial distance but also fluctuations in the radial direction.

The left panel of Figure~\ref{fig:3d_electron_in_h_atom} shows the trajectory of the electron in the $(2,0,0)$ state. The cutaway exposes the interior of the cloud: the electron samples both the inner core and the outer shell of the state, consistent with the radial distribution shown in Figure~\ref{fig:radial_distribution_2s0}.

In the right panel of Figure~\ref{fig:3d_electron_in_h_atom}, we present a simulation of the $(2,1,0)$ state of the hydrogen atom, generated 2~\text{ps} after the simulation commenced. The electron is observed to traverse both lobes of the $(2,1,0)$ state. In contrast, shorter simulation durations confine the electron to a single lobe, as evidenced by the polar distribution of the particle’s position depicted in Figure~\ref{fig:angular_distribution_2p0}.

\section{Radial, Polar, and Azimuthal Distributions of Electron Positions Around the Atom}
\label{sec:coordinate-distributions}
\begin{figure}[ht]
  \vspace{0pt}
    \centering
    \begin{minipage}{0.32\textwidth}
        \centering
        \includegraphics[width=\textwidth]{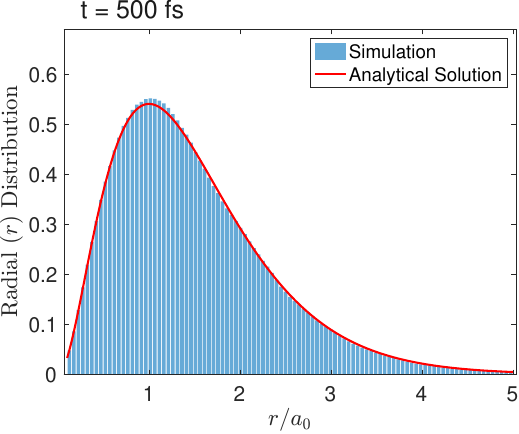}
    \end{minipage}
    \begin{minipage}{0.32\textwidth}
        \centering
        \includegraphics[width=\textwidth]{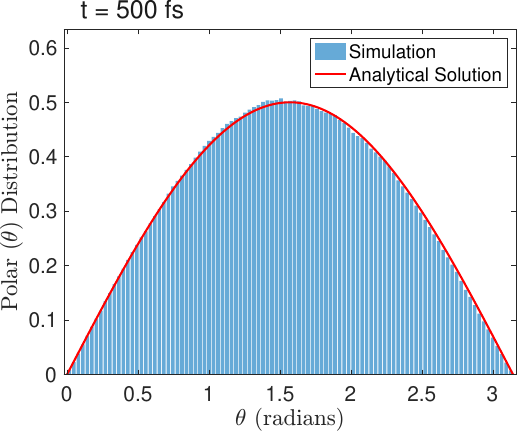}
    \end{minipage}
    \begin{minipage}{0.32\textwidth}
        \centering
        \includegraphics[width=\textwidth]{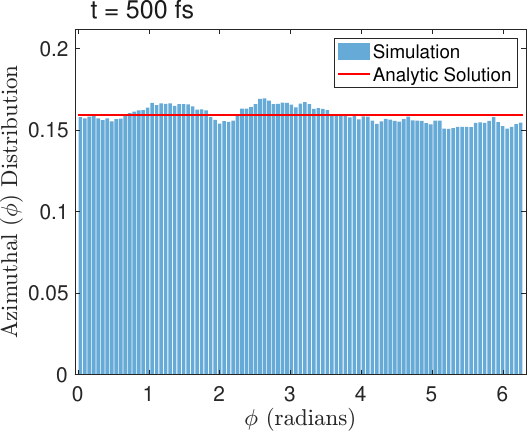}
    \end{minipage}
      \vspace{0pt}
    \caption{Radial, polar, and azimuthal distributions of the position of the electron, simulated as a single particle moving in the forward drift field $\mathbf b = \frac{\hbar}{m_e} \, \nabla \log |\psi_{1,0,0}|$. The simulation results show good agreement with the analytical result for the normalized probability distribution.}
    \label{fig:radial_polar_azimuthal_distribution}
\end{figure}

We record the electron’s spherical coordinates along the simulated trajectory and histogram each coordinate component. Figure~\ref{fig:radial_polar_azimuthal_distribution} shows that, after sufficient time, the radial, polar, and azimuthal distributions of the electron coordinates in the $(1,0,0)$ state match the theoretical distributions predicted by the Born rule and the theoretical wave function.

If $R_{n\ell}(r)$ is the radial part of the wave function defined in Eq.~\eqref{eq:radial_wave_function}, $\Theta_{\ell m}(\theta)$ is the polar part defined in Eq.~\eqref{eq:normalized_legendre_poly}, and $\Phi_{m}(\phi)$ is the azimuthal part defined in Eq.~\eqref{eq:phi_m_complex}, the probability distribution of the electron’s spherical components is:
\begin{equation}
\label{eq:probability_distributions}
\begin{aligned}
P_r(r) & = r^2 (R_{n\ell}(r))^2, \\
P_\theta(\theta) & = 2 \pi \sin \theta \, (\Theta_{\ell m}(\theta))^2, \\
P_\phi(\phi) & = \frac{1}{2 \pi}|\Phi_{m}(\phi)|^2 = \frac{1}{2 \pi}.
\end{aligned}
\end{equation}

\begin{figure}[h!]
  \vspace{0pt}
    \centering
    \begin{minipage}{0.32\textwidth}
        \centering
        \includegraphics[width=\textwidth]{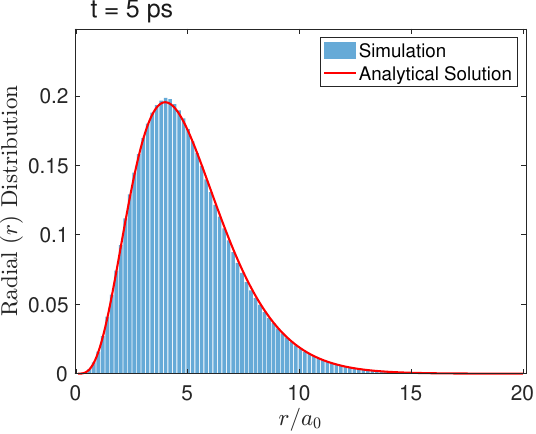}
    \end{minipage}\hspace{4pt}
    \begin{minipage}{0.32\textwidth}
        \centering
        \includegraphics[width=\textwidth]{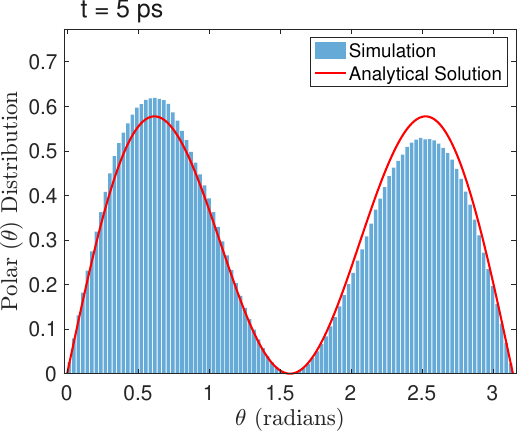}
    \end{minipage}\hspace{4pt}
    \begin{minipage}{0.32\textwidth}
        \centering
        \includegraphics[width=\textwidth]{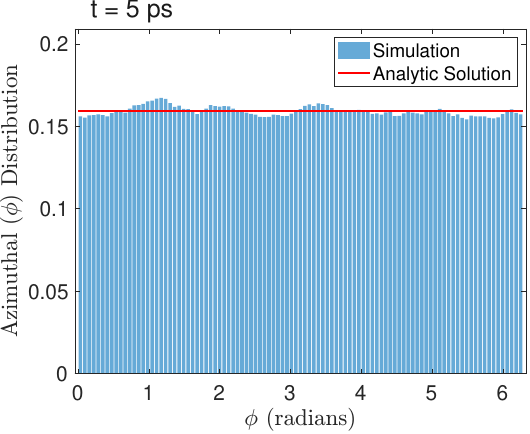}
    \end{minipage}
      \vspace{0pt}
    \caption{Radial and angular distributions of the position of the electron, simulated as a single particle moving in the forward drift field $\mathbf b= \frac{\hbar}{m_e} \,\nabla \log |\psi_{2,1,0}|$. The simulation results show good agreement with the analytical result for the normalized probability distribution.}
    \label{fig:radial_and_azimuthal_distribution_2p0}
\end{figure}

In Figure~\ref{fig:radial_and_azimuthal_distribution_2p0}, we show the radial and angular distributions of the electron in the $(2,1,0)$ state after a sufficiently long simulation time. The simulation results are in good agreement with the theoretical probability densities.

\begin{figure}[h!]
  \vspace{0pt}
    \centering
    \begin{minipage}{0.32\textwidth}
        \centering
        \includegraphics[width=\textwidth]{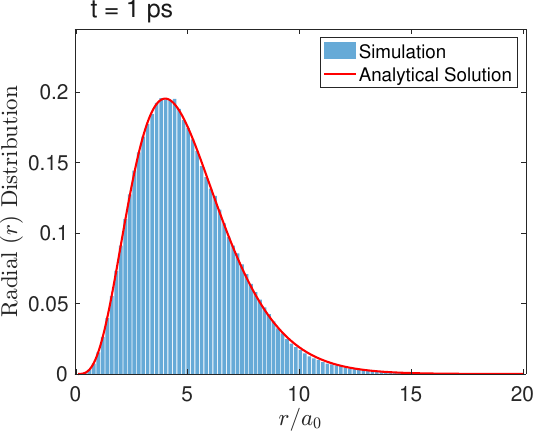}
    \end{minipage}\hspace{4pt}
    \begin{minipage}{0.32\textwidth}
        \centering
        \includegraphics[width=\textwidth]{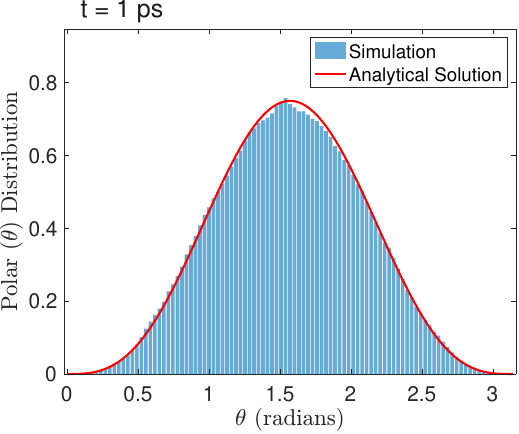}
    \end{minipage}\hspace{4pt}
    \begin{minipage}{0.32\textwidth}
        \centering
        \includegraphics[width=\textwidth]{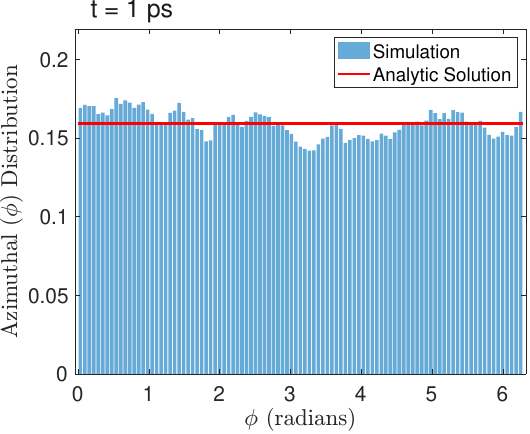}
    \end{minipage}
      \vspace{0pt}
    \caption{Radial and angular distributions of the position of the electron, simulated as a single particle moving in the forward drift field~\(\mathbf b=\frac{\hbar}{m_e}\,\nabla\log|\psi_{2,1,1}|
    +\frac{\hbar}{m_e}\,\nabla\operatorname{Arg}\psi_{2,1,1}\). The
    simulation results show good agreement
    with the analytical result for the normalized probability distribution.}
    \label{fig:distributions_2p1}
\end{figure}

The corresponding distributions for the \((2,1,1)\) state are shown in
Figure~\ref{fig:distributions_2p1}. Its radial distribution coincides with
that of the \((2,1,0)\) state, the polar density
\(P_\theta=\frac{3}{4}\sin^3\theta\) concentrates at the equatorial plane
instead of the poles, and the azimuthal distribution remains uniform even
though the drift field now contains a non-vanishing azimuthal component.
The dynamical role of this phase-driven component is analyzed in
Section~\ref{subsec:azimuthal-current}.

\begin{figure}[h!]
  \vspace{0pt}
    \centering
    \begin{minipage}{0.32\textwidth}
        \centering
        \includegraphics[width=\textwidth]{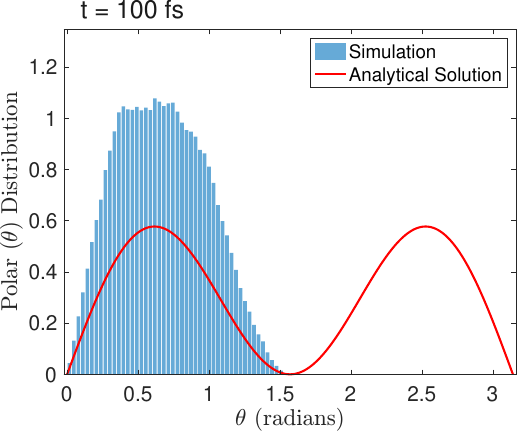}
    \end{minipage}\hspace{4pt}
    \begin{minipage}{0.32\textwidth}
        \centering
        \includegraphics[width=\textwidth]{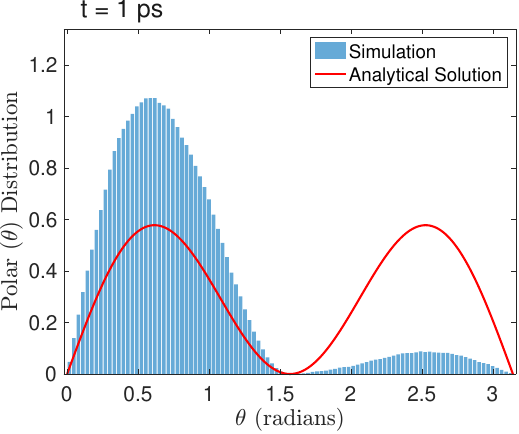}
    \end{minipage}\hspace{4pt}
    \begin{minipage}{0.32\textwidth}
        \centering
        \includegraphics[width=\textwidth]{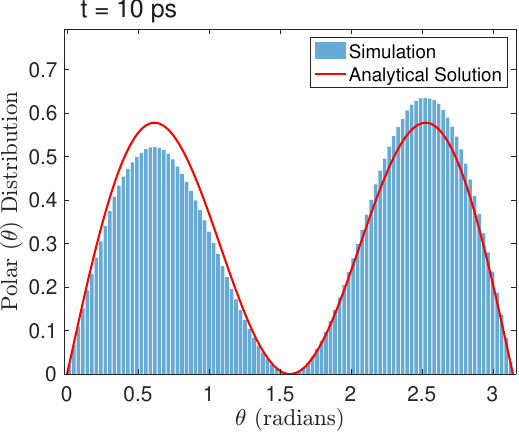}
    \end{minipage}
      \vspace{0pt}
   \caption{Polar distributions of the electron's position, simulated as a single particle moving in the forward drift field $\mathbf{b} = \frac{\hbar}{m_e} \, \nabla \log |\psi_{2,1,0}|$. After sufficient time, the simulation results show good agreement with the analytical probability distribution. The three panels show the distribution accumulated up to increasing simulation times, from left to right.}
    \label{fig:angular_distribution_2p0}
\end{figure}

\begin{figure}[h!]
  \vspace{0pt}
    \centering
    \begin{minipage}{0.32\textwidth}
        \centering
        \includegraphics[width=\textwidth]{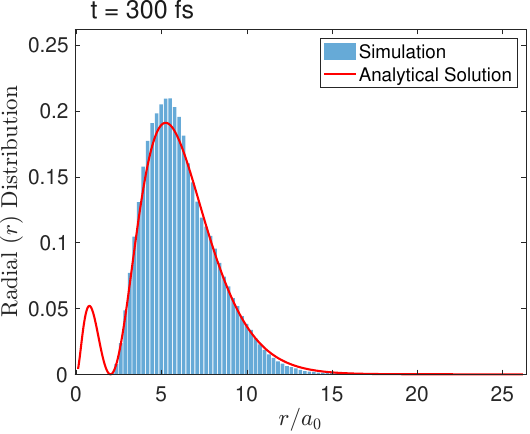}
    \end{minipage}\hspace{4pt}
    \begin{minipage}{0.32\textwidth}
        \centering
        \includegraphics[width=\textwidth]{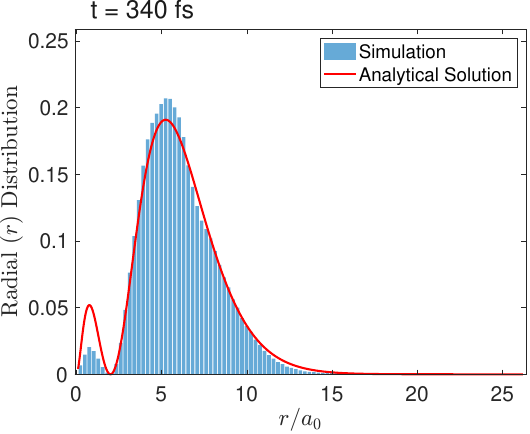}
    \end{minipage}\hspace{4pt}
    \begin{minipage}{0.32\textwidth}
        \centering
        \includegraphics[width=\textwidth]{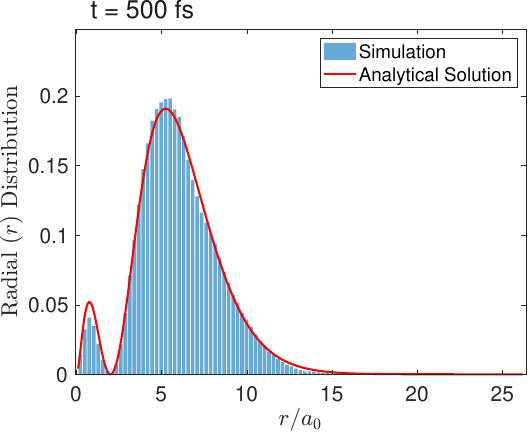}
    \end{minipage}
      \vspace{0pt}
   \caption{Radial distributions of the electron's position, simulated as a single particle moving in the forward drift field $\mathbf{b} = \frac{\hbar}{m_e} \, \nabla \log |\psi_{2,0,0}|$. After sufficient time, the simulation results show good agreement with the analytical probability distribution. The three panels show the distribution accumulated up to increasing simulation times, from left to right.}
    \label{fig:radial_distribution_2s0}
\end{figure}

The polar distribution of the $(2,1,0)$ state shown on the right side of Figure~\ref{fig:angular_distribution_2p0} follows the theoretical prediction after a sufficiently long simulation time. However, for shorter simulation times, the electron becomes trapped in one of the two lobes, with very rare jumps to the other lobe, as seen in the left side of Figure~\ref{fig:angular_distribution_2p0}. During the simulation, it is typical to observe an asymmetric polar electron distribution, which becomes symmetric after sufficient simulation time.

The radial distribution of the $(2,0,0)$ state shown on the right side of Figure~\ref{fig:radial_distribution_2s0} follows the theoretical prediction after a sufficiently long simulation time. Similar to the angular distribution of the $(2,1,0)$ state, for shorter simulation times, the electron becomes trapped in one of the two shells, with very rare jumps to the other shell, as seen in the left side of Figure~\ref{fig:radial_distribution_2s0}.

\begin{figure}[ht]
  \vspace{0pt}
    \centering
    \begin{minipage}{0.40\textwidth}
        \centering
        \includegraphics[width=\textwidth]{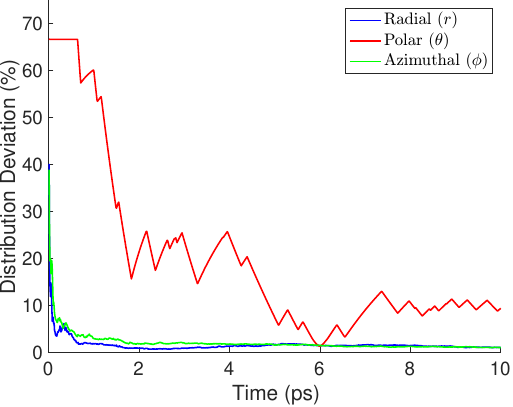}
    \end{minipage}\hspace{30pt}
    \begin{minipage}{0.40\textwidth}
        \centering
       \includegraphics[width=\textwidth]{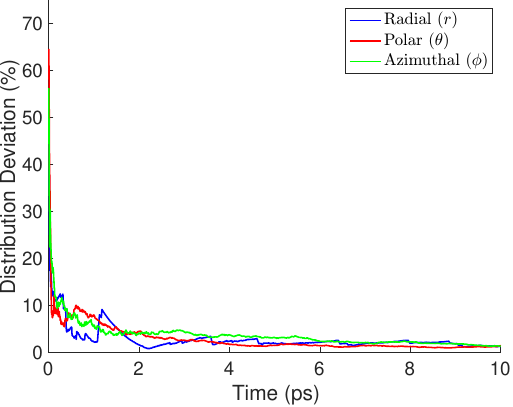}
    \end{minipage}
      \vspace{0pt}
   \caption{Deviations of the simulated radial and angular distributions from the theoretical distributions predicted by the corresponding wave functions and Born rule. In the left figure, the shown state is the $(2,1,0)$ state, and in the right figure, the shown state is the $(2,0,0)$ state. After the initial few femtoseconds most distributions show only minor deviations. The exception in each state is the coordinate split in two by a node, which converges only over a few picoseconds: the polar distribution of the $(2,1,0)$ state and the radial distribution of the $(2,0,0)$ state. The ripples in these two distributions are attributed to rare transitions of the electron across the node, between the two lobes of the $(2,1,0)$ state and between the two shells of the $(2,0,0)$ state.}
    \label{fig:distributions_deviation_2p0}
\end{figure}

In Figure~\ref{fig:distributions_deviation_2p0}, we show the deviations of the radial, polar, and azimuthal distributions of the electron's positions from the theoretical probability distributions predicted by the wave function and the Born rule (see Eq.~\eqref{eq:probability_distributions}). One can see that for the $(2,1,0)$ and $(2,0,0)$ states, after a sufficiently long time, the deviations of all distributions become smaller and smaller. This means that the wave function, through the Born rule, reproduces the statistics of the single electron's position only after sufficient averaging over the trajectory.

The eigenstate density $|\psi_{n\ell m}|^2$ is the stationary (invariant) distribution of the stochastic dynamics, not a quantity that evolves in time. The relevant times are therefore the ergodic mixing times of a single trajectory, the time its empirical distribution needs to converge to $|\psi|^2$, governed by the relaxation spectrum of the Fokker--Planck operator, which becomes slow when a nodal surface separates the support into metastable regions. For the $(2,1,0)$ state the radial and azimuthal distributions settle within a few femtoseconds, while the polar distribution requires a few picoseconds because of the rare transitions between the two lobes. For the $(2,0,0)$ state the roles are reversed: the radial distribution is the slow one, converging over a few picoseconds through rare jumps between the two shells.

\subsection{Phase-driven azimuthal circulation for the $m=\pm1$ states}
\label{subsec:azimuthal-current}

For the \(m=0\) states the spatial phase of the wave function is
constant, and the optimal drift is purely osmotic,
\(\mathbf b=\frac{\hbar}{m_e}\nabla\log|\psi|\). The magnetic eigenstates with
\(m\neq0\) provide the complementary test of the formulation: as shown at
the end of Section~\ref{sec:wavefunction-drift-hydrogen}, their azimuthal
drift is generated entirely by the phase factor
\(\Phi_m(\phi)=e^{im\phi}\) and coincides with the current velocity,
\(b_\phi=v_\phi=\hbar m/(m_e r\sin\theta)\).

Two quantitative signatures follow from the Born averages of the
\((2,1,\pm1)\) states, using the factorized form of the Born density.

First, the mean azimuthal velocity is
\begin{equation}
\label{eq:mean_azimuthal_velocity_2p1}
\langle b_\phi\rangle_{2,1,\pm1}
=
\frac{\hbar m}{m_e}
\left\langle\frac{1}{r}\right\rangle_{2,1}
\left\langle\frac{1}{\sin\theta}\right\rangle_{1,\pm1}
=
\frac{\hbar m}{m_e}\,
\frac{1}{4a_0}\cdot\frac{3\pi}{8}
=
m\,\frac{3\pi}{32}\,\frac{\hbar}{m_e a_0}
\approx
m\times6.44\times10^{5}\ \mathrm{m/s},
\end{equation}
where the subscripts record the quantum numbers each average depends on. The
radial average depends only on \((n,\ell)\), so for the \((2,1,\pm1)\) states
\(\langle 1/r\rangle_{2,1}=1/(4a_0)\), while the polar average depends only on
\((\ell,m)\), giving
\(\langle 1/\sin\theta\rangle_{1,\pm1}
=2\pi\!\int_0^\pi\left|\Theta_{1,\pm1}(\theta)\right|^2 d\theta=3\pi/8\).

Second, the unwrapped azimuthal coordinate winds ballistically: from the
azimuthal equation of motion
Eq.~\eqref{eq:final_stochastic_equation_of_motion}, the mean winding rate is
\begin{equation}
\label{eq:winding_rate_2p1}
\bigl\langle\dot\phi\bigr\rangle_{2,1,\pm1}
=
\frac{\hbar m}{m_e}
\left\langle\frac{1}{r^2}\right\rangle_{2,1}
\left\langle\frac{1}{\sin^2\theta}\right\rangle_{1,\pm1}
=
\frac{\hbar m}{m_e}\,
\frac{1}{12a_0^2}\cdot\frac{3}{2}
=
\frac{m\hbar}{8m_ea_0^2}
\approx
m\times5.17\ \mathrm{rad/fs},
\end{equation}
about \(0.8\) revolutions per femtosecond, whereas for \(m=0\) the azimuth
performs an unbiased random walk.

Moreover, the form of \(b_\phi\) in
Eq.~\eqref{eq:drift_velocity_spherical} makes the \(z\)-component of the
orbital angular momentum carried by the drift an exact pointwise constant
of the motion,
\begin{equation}
\label{eq:Lz_quantization_trajectory}
L_z
=
m_e\,r\sin\theta\,b_\phi
=
m\hbar,
\end{equation}
along every trajectory, not merely in the Born-density average: the quantization of \(L_z\)
postulated in the Bohr--Sommerfeld model reappears here as a deterministic
property of the stochastic motion.
The azimuthal drift coincides with the de Broglie--Bohm guidance
velocity~\cite{Bohm1952a,Bohm1952b}, for which the eigenstate trajectories
are circles of constant \(L_z\), each sampling only its own orbit.
In the present stochastic dynamics, by
contrast, a single trajectory of the ergodic dynamics samples the full Born
density while its drift continues to carry exactly \(L_z=m\hbar\).

\begin{figure}[ht]
  \centering
  \includegraphics[width=\textwidth]{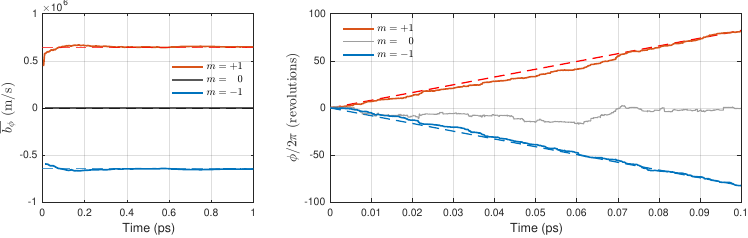}
  \caption{Phase-driven azimuthal circulation of the electron in the
    \((2,1,m)\) states, \(m=+1,0,-1\), simulated as single real-valued
    stochastic particles with the forward drift field
    \(\mathbf b=\frac{\hbar}{m_e}\nabla\log|\psi_{2,1,m}|
    +\frac{\hbar}{m_e}\nabla\operatorname{Arg}\psi_{2,1,m}\).
    \textbf{Left:} Running time-average of the azimuthal drift \(b_\phi\).
    The dashed lines mark the analytical values of
    Eq.~\eqref{eq:mean_azimuthal_velocity_2p1} and zero.
    \textbf{Right:} Unwrapped azimuthal coordinate, shown as the number of
    full revolutions \(\phi/2\pi\) over the first \(100\)~fs. The dashed
    lines mark the analytical winding rates of
    Eq.~\eqref{eq:winding_rate_2p1}.}
  \label{fig:azimuthal_current_2p1}
\end{figure}

Figure~\ref{fig:azimuthal_current_2p1} shows these signatures for the states
\((2,1,+1)\), \((2,1,0)\), and \((2,1,-1)\): the running time-average of
\(b_\phi\) converges to the value in Eq.~\eqref{eq:mean_azimuthal_velocity_2p1} with
the corresponding sign and vanishes identically for \(m=0\) (left panel),
while the unwrapped azimuth grows linearly at the
rate in Eq.~\eqref{eq:winding_rate_2p1} for \(m=\pm1\) and only diffuses for
\(m=0\) (right panel). The coordinate distributions of these trajectories,
shown in Figure~\ref{fig:distributions_2p1}, are identical for the two
circulating states: the circulation is therefore invisible in the static
Born density but fully resolved at the level of the single stochastic
trajectory, where the phase \(S\) of the reconstructed wave function acts
through the optimal drift, as built into
Eq.~\eqref{eq:forward_drift_wavefunction}.

Equation~\eqref{eq:Lz_quantization_trajectory} is a property of the drift
field and presupposes the drift law. To recover the quantization from the
simulated motion itself, using only kinematic trajectory data, consider the
running time-average of the angular momentum accumulated along the
trajectory,
\begin{equation}
\label{eq:Lz_running_estimator}
\bar L_z(t)
=
\frac{1}{t}\int_0^t m_e\,r^2\sin^2\!\theta\;d\phi,
\end{equation}
computed in the It\^o sense, with \(d\phi\) the azimuthal increment actually
applied by the integrator, drift and noise alike, so the drift formula does
not enter this quantity.
Here and in what follows, an overbar denotes a running time average along a
single trajectory, whereas angle brackets \(\langle\cdot\rangle\) denote
averages over the Born density.
As shown in Appendix~\ref{app:Lz_convergence}, substituting the azimuthal
equation of motion~\eqref{eq:final_stochastic_equation_of_motion} splits
\(\bar L_z(t)\) into the constant drift contribution \(m\hbar\) of
Eq.~\eqref{eq:Lz_quantization_trajectory} and a zero-mean martingale, the
Brownian part of the azimuthal increment that the estimator retains. The
running average is therefore an unbiased estimate of the quantized value at
every \(t\), independently of the radial and angular profile of the state.
Its deviations from \(m\hbar\) are this martingale, which averages out over
time with the \(t^{-1/2}\) fluctuation envelope
\begin{equation}
\label{eq:Lz_std}
\left.\frac{\Delta\bar L_z(t)}{\hbar}\right|_{2,1,\pm1}
=
\sqrt{\frac{24\,m_e a_0^2}{\hbar\,t}}
\approx
2.4\%
\quad
\text{at } t=1~\mathrm{ps},
\end{equation}
where \(\Delta\bar L_z(t)\) denotes the standard deviation of
\(\bar L_z(t)\) over realizations of the noise, and the subscript denotes
the state whose Born averages enter the prefactor.

\begin{figure}[ht]
  \centering
  \includegraphics[width=0.45\textwidth]{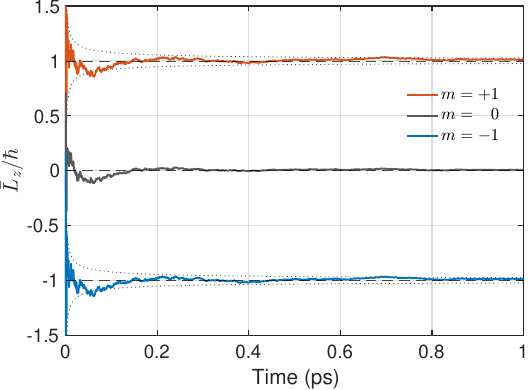}
  \caption{Running time-average \(\bar L_z(t)\) of the angular momentum
    accumulated along the trajectory, Eq.~\eqref{eq:Lz_running_estimator},
    in units of \(\hbar\), computed from the raw azimuthal increments of
    single simulated trajectories of the \((2,1,m)\) states, \(m=+1,0,-1\).
    The dashed lines mark \(+1\), \(0\), and \(-1\). The thin dotted
    curves around \(\pm1\) mark the analytical one-standard-deviation
    envelope \(m\pm\Delta\bar L_z(t)/\hbar\) of Eq.~\eqref{eq:Lz_std}.}
  \label{fig:lz_running_estimator}
\end{figure}

Figure~\ref{fig:lz_running_estimator} shows \(\bar L_z(t)/\hbar\) for the
three states \((2,1,+1)\), \((2,1,0)\), and \((2,1,-1)\): the running
average converges to \(+1\), \(0\), and \(-1\), respectively, with
fluctuations of the size set by the analytical
envelope~\eqref{eq:Lz_std}. The quantization of \(L_z\) is
therefore not only an analytical identity of the optimal drift field but is
recovered numerically from the raw increments of the simulated stochastic
motion.

\section{Kinetic, Potential, and Total Energy of the Electron}
\label{sec:electron-energies}
By obtaining the forward-drift components and position of the electron at each simulation time step, we can easily compute the radial ($T_r$), polar ($T_\theta$), and azimuthal ($T_\phi$) kinetic energies, as well as the potential energy ($V$) using simple classical equations of particle energies, as follows:
\begin{equation}
T_r = \frac{1}{2} m_e b_r^2, \qquad
T_\theta = \frac{1}{2} m_e b_\theta^2, \qquad
T_\phi = \frac{1}{2} m_e b_\phi^2, \qquad
V= - \frac{e^2}{4 \pi \epsilon_0 r}.
\end{equation}

The total energy is just the sum of all energies:
\begin{equation}
E = T_r + T_\theta + T_\phi + V.
\end{equation}

In this section, we present the results for the average electron energies in the $(2,1,0)$ and $(2,0,0)$ states of the hydrogen atom. Similar results can be obtained for any other state by running the provided code~\cite{github_Yordanov2024} with different quantum numbers.

\begin{figure}[ht]
  \vspace{0pt}
    \centering
    \begin{minipage}{0.40\textwidth}
        \centering
        \includegraphics[width=\textwidth]{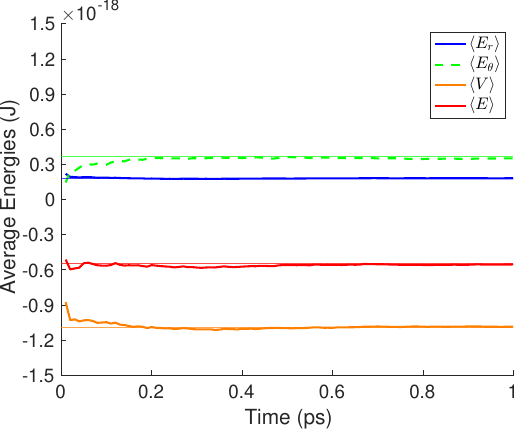}
    \end{minipage}\hspace{30pt}
    \begin{minipage}{0.40\textwidth}
        \centering
       \includegraphics[width=\textwidth]{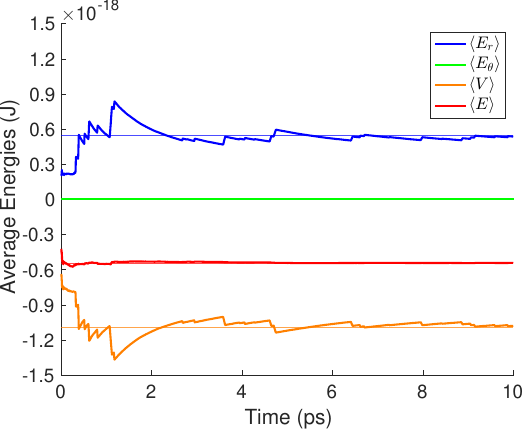}
    \end{minipage}
      \vspace{0pt}
   \caption{Average radial and angular kinetic energies, along with potential and total energies, of the electron under stochastic evolution with forward drift fields defined by $\mathbf{b}_{2,1,0} = \frac{\hbar}{m_e} \, \nabla \log |\psi_{2,1,0}|$ and $\mathbf{b}_{2,0,0} = \frac{\hbar}{m_e} \, \nabla \log |\psi_{2,0,0}|$, plotted as functions of simulation time for the $(2,1,0)$ (left) and $(2,0,0)$ (right) states. After a few hundred femtoseconds for the $(2,1,0)$ state and a few picoseconds for the $(2,0,0)$ state, the simulated energies are observed to converge to their theoretical expectations. The results shown are for a single stochastic trajectory ($M = 1$).}
   \label{fig:energies_2p0_2s0}
\end{figure}

The results for the average energies as a function of simulation time are shown in Figure~\ref{fig:energies_2p0_2s0}.

The average radial energy for the chosen atomic state can be calculated using Eq.~\eqref{eq:radial_KE_stochastic} and the radial wave function~\eqref{eq:radial_wave_function}, applied to the states $(2,1,0)$ and $(2,0,0)$ in our example.
\begin{equation}
\label{eq:avg_T_r_2p0_2s0}
\langle T_r \rangle_{2,1,0} = 1.8166 \times 10^{-19} \, \text{J}, \qquad \langle T_r \rangle_{2,0,0} = 5.4497 \times 10^{-19} \, \text{J}.
\end{equation}

In order to calculate the average angular energy, it is necessary first to calculate the average of the reciprocal square of the radial position of the electron for the $(2,1,0)$ state:
\begin{equation}
\label{eq:avg_1_r^2_2p0}
\left< \frac{1}{r^2} \right>_{2,1,0} = \int_0^\infty r^2 \left[ R_{21}(r) \right]^2 \frac{1}{r^2} \, dr =  \frac{1}{12 a_0^2} = 2.9759 \times 10^{19} \, \text{m}^{-2}.
\end{equation}

The average angular energy for the chosen atomic states can be calculated using Eqs.~\eqref{eq:avg_angular_kinetic_energy_final} and~\eqref{eq:avg_1_r^2_2p0}:
\begin{equation}
\label{eq:avg_T_angular_2p0_2s0}
\langle T_{\text{angular}} \rangle_{2,1,0} = 3.6331 \times 10^{-19} \,\text{J}, \qquad \langle T_{\text{angular}} \rangle_{2,0,0} = 0\,\text{J}.
\end{equation}

The potential energy of the $(2,1,0)$ and $(2,0,0)$ states can be calculated with the following integral:
\begin{equation}
\label{eq:avg_V_2p0_2s0}
\begin{aligned}
\left< V \right>_{2,1,0}  &= - \frac{e^2}{4 \pi \epsilon_0} \left< \frac{1}{r} \right>_{2,1,0} = - \frac{e^2}{4 \pi \epsilon_0} \int_0^\infty r^2 \left[ R_{21}(r) \right]^2 \frac{1}{r} \, dr \\
&=  - \frac{e^2}{4 \pi \epsilon_0} \frac{1}{4 a_0} = - 10.8994 \times 10^{-19} \, \text{J},\\
\left< V \right>_{2,0,0}  &= -10.8994 \times 10^{-19}\, \text{J}.
\end{aligned}
\end{equation}

Finally, the total energies of the electron for the $(2,1,0)$ and $(2,0,0)$ states are:
\begin{equation}
\label{eq:avg_E_2p0}
\begin{aligned}
\left< E \right>_{2,1,0} &= \langle T_r \rangle_{2,1,0} + \langle T_{\text{angular}} \rangle_{2,1,0}  + \left< V \right>_{2,1,0} = -5.4497 \times 10^{-19} \,\text{J}, \\
\left< E \right>_{2,0,0} &=  -5.4497 \times 10^{-19} \,\text{J}.
\end{aligned}
\end{equation}

It can be seen from Figure~\ref{fig:energies_2p0_2s0} that the calculated values for the average radial and angular kinetic energies, potential energy, and total energy in Eqs.~\eqref{eq:avg_T_r_2p0_2s0}, \eqref{eq:avg_T_angular_2p0_2s0}, \eqref{eq:avg_V_2p0_2s0}, and~\eqref{eq:avg_E_2p0} are in good agreement with the corresponding energies of the electron computed from the computer simulation after sufficient time.

\section{Analysis of Stability and Convergence in Stochastic Simulations}
\label{sec:simulation_convergence_2p0}

In Section~\ref{sec:electron-energies} the simulated energies of the
$(2,1,0)$ and $(2,0,0)$ states were found to agree with the analytical values
only after a sufficiently long simulation time. We now examine what controls
this convergence. Both states are nodal. The wave function $\psi_{2,1,0}$
vanishes on the polar plane $\theta=\pi/2$ between its two lobes, and
$\psi_{2,0,0}$ on the radial sphere $r=2a_0$ between its two shells
(Figures~\ref{fig:angular_distribution_2p0}
and~\ref{fig:radial_distribution_2s0}). On these nodal sets the
drift is singular (see Eq.~\eqref{eq:drift_velocity_spherical}), as already noted in
Section~\ref{sec:hydrogen-simulation}. This singularity is what makes the
convergence slow. For these two nodal states the kinetic-energy distributions
are heavy-tailed, as derived in
Appendix~\ref{app:analytical_kinetic_energy_distributions}. At fixed $r$, the
polar kinetic energy $T_\theta$ of $(2,1,0)$ follows an exact Lomax
distribution. The radial kinetic energy $T_r$ of $(2,0,0)$ has a Lomax-type
power-law tail. In both cases the probability density decays as $T^{-5/2}$ at
large energies, so the mean is finite but the variance is infinite. The
large kinetic energies responsible for the divergent variance occur when the
trajectory passes very close to the node, where the drift diverges.

Because the variance is infinite, the mean kinetic energy would converge only
over an infinitely long simulation time. To make the average converge within a
finite, practical computational time, we cut the heavy tail by introducing a
maximum cutoff velocity $v_{\max}$: in the simulation each drift term entering
the equations of motion~\eqref{eq:final_stochastic_equation_of_motion} is
constrained to remain below this value, which bounds the variance and restores
finite-time convergence. We then investigate how the average kinetic energy
depends on the velocity cutoff $v_{\max}$ and the discretization time step
$\Delta t$. The aim of this study is to establish a criterion for choosing the
values of these parameters such that they have minimal influence on the physical
result.

In addition to the velocity cutoff, we regularize the coordinate singularities
at the origin and poles by keeping $r$ and $\sin\theta$ bounded away from zero.
Both the drift and the stochastic terms of the stochastic equation of motion~\eqref{eq:final_stochastic_equation_of_motion} contain $1/r$ and
$1/\sin\theta$, so this coordinate bound in general affects both. However, where
this bound acts the drift is already limited by the velocity cutoff, so the
coordinate bound modifies only the stochastic terms $(\sigma/r)\,dW^\theta$ and
$(\sigma/(r\sin\theta))\,dW^\phi$, i.e.\ the spatial sampling near the origin and
poles; it does not change the kinetic energy, which is built from the
already-capped drift. The implementation additionally floors
\(|R_{n\ell}|\) and \(|\Theta_{\ell m}|\) in the log-derivatives of
Eq.~\eqref{eq:final_stochastic_equation_of_motion} to avoid division by zero
exactly on the nodal sets. Wherever this floor acts, the resulting drift far
exceeds \(v_{\max}\), so the drift actually applied is determined by the
velocity cutoff alone.

\begin{figure}[ht]
  \centering
  \begin{minipage}{0.49\textwidth}
    \centering
    \includegraphics[width=\textwidth]{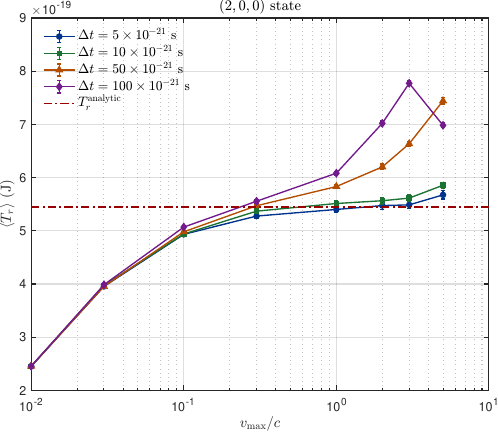}
  \end{minipage}\hfill
  \begin{minipage}{0.49\textwidth}
    \centering
    \includegraphics[width=\textwidth]{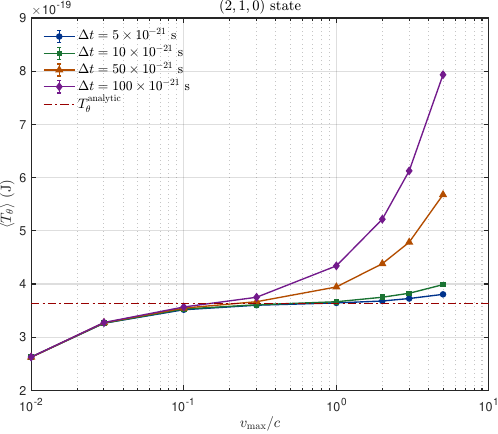}
  \end{minipage}
  \caption{Computed kinetic energy versus the velocity cutoff $v_{\max}/c$ for
    the two nodal states, at different integration time steps $\Delta t$.
    \textbf{Left:} radial kinetic energy $\langle T_r \rangle$ for the
    $(2,0,0)$ state, with analytical value
    $T^{\mathrm{analytic}}_r = 5.450 \times 10^{-19}$\,J. \textbf{Right:} polar
    kinetic energy $\langle T_\theta \rangle$ for the $(2,1,0)$ state, with
    $T^{\mathrm{analytic}}_\theta = 3.633 \times 10^{-19}$\,J. In both panels
    $M = 100$ trajectories are used, and the error bars are bootstrap
    standard errors of the mean.
    The dash-dotted line is the analytical value, shown for comparison
    and not used in selecting $\Delta t$ or $v_{\max}$. Refining
    $\Delta t$ drives the average onto a plateau on which it is insensitive to
    both $\Delta t$ and $v_{\max}$; the velocity cutoff serves only to make the
    average converge in a finite simulation time. The converged estimate is read
    on this plateau.}
  \label{fig:cutoff_scan_2s0}
\end{figure}

We select the time step and the cutoff from the internal convergence of the
simulation. The criterion is the collapse of the curves for different
$\Delta t$ onto a common plateau. This selection does not use the analytical
value, which is overlaid in Figure~\ref{fig:cutoff_scan_2s0} only for
comparison. Sweeping $\Delta t$
and $v_{\max}$ (Figure~\ref{fig:cutoff_scan_2s0}) shows that refining the time
step drives $\langle T_r \rangle$ toward a limiting value: at coarse $\Delta t$
the average lies above it, and as $\Delta t$ is reduced the result decreases and
settles. For $v_{\max} \gtrsim c$ this limit is reached---the curves for
successively smaller $\Delta t$ converge and flatten into a plateau on which
$\langle T_r \rangle$ is insensitive to both $\Delta t$ and $v_{\max}$. For
$v_{\max} \ll c$ the cutoff is too restrictive: the result lies well below the
plateau and no longer responds to $\Delta t$, so the plateau has a lower edge
near $v_{\max} \approx c$. Each $\langle T_r \rangle$ is reported with a nonparametric bootstrap standard
error of the cutoff-regularized distribution, which quantifies the statistical
uncertainty under the heavy tail.

This plateau is the operational convergence criterion. The converged estimate is the plateau value, read at the largest
$\Delta t$ that has settled onto the plateau and over the range of $v_{\max}$
for which $\langle T_r \rangle$ no longer changes within its statistical
scatter.

The total simulated time, $T = 20\,\mathrm{ps}$ per run, the same for both
states and every $\Delta t$, is
long enough to reach this plateau. The running time averages
$\bar T_r(T/2)$ and $\bar T_r(T)$ agree
within the statistical resolution, confirming the average has stopped drifting. The residual rise at the coarsest steps is a discretization artifact that
vanishes as $\Delta t \to 0$, not a physical effect. On the plateau the velocity cutoff
engages on $\lesssim 10^{-7}$ of the integration steps, so its effect on the
dynamics is negligible. There $\langle T_r \rangle$ agrees with the analytical
value to $\approx 1$--$2\%$, consistent with the heavy-tail statistical
uncertainty.

The $(2,1,0)$ state behaves identically, with $\langle T_\theta \rangle$ in
place of $\langle T_r \rangle$ (Figure~\ref{fig:cutoff_scan_2s0}, right): its
polar kinetic energy has the same Lomax-type tail
(Appendix~\ref{app:analytical_kinetic_energy_distributions}), and refining
$\Delta t$ produces a plateau near $v_{\max}\approx c$ that agrees with the
analytical value
$T^{\mathrm{analytic}}_\theta = 3.633\times 10^{-19}\,\mathrm{J}$. For both
nodal states the plateau criterion thus recovers the correct kinetic energy,
despite the drift singularity at the node.

We close by distinguishing this slow convergence from genuine non-ergodicity.
In the strict nonrelativistic idealization the drift
field in Eq.~\eqref{eq:drift_velocity_spherical} diverges on the nodal set, which
therefore acts as an impenetrable boundary for the idealized process. The
divergence, however, is reached only inside a thin shell around the nodes: the
drift exceeds the speed of light within a reduced Compton wavelength
\(\hbar/(m_e c)\approx 0.007\,a_0\) of the nodal set, precisely where the
nonrelativistic description has already ceased to be physically meaningful.
The velocity cutoff \(v_{\max}\sim c\) is therefore a physical regularization
of the idealized drift law, not merely a numerical convenience. With the
cutoff in place the process is an ordinary ergodic Markov diffusion driven by
Wiener noise: its time averages converge to the stationary ensemble averages
of the regularized dynamics, whose stationary density agrees with the Born
density $P=|\psi|^2$ up to a correction localized near the nodal set and
controlled by $v_{\max}$. On the plateau of
Figure~\ref{fig:cutoff_scan_2s0} this correction is below the statistical
resolution of the computed averages and distributions. The slow convergence of
the energy averages is then the familiar slowness of averaging an
infinite-variance quantity, which the cutoff cures by making the variance
finite. This is distinct from
anomalous-diffusion models such as fractional Brownian motion with nonlinear
clocks or under confinement, which can age and remain genuinely non-ergodic
even in bounded domains \cite{Liang2023PRE034113,Liang2023PREL052101}, so that
time and ensemble averages may differ even over the longest experimentally or
numerically accessible time windows. Here, by contrast, the two averages
coincide for the regularized dynamics; the only price of the heavy tail is the
rate of convergence, which the cutoff controls.


\section{Conclusion}
\label{sec:conclusion}

We have formulated the quantum dynamics of a single electron as a
law-dependent stochastic optimal-control problem in which the forward drift of
the stochastic process is the control field. Once the forward drift is
specified, the backward drift of the same process is fixed by its density
through the forward--backward time-reversal relation. Consequently, the chosen
bilinear kinetic term makes the running cost depend on the evolving probability
density. The control problem is therefore posed on density space, with \(P\) as
the state variable and \(\mathcal V(t,P)\) as the Bellman value functional. The
evolution of \(P\) is governed by the corresponding forward Fokker--Planck
equation, which serves as the state equation. Bellman's principle yields the
functional HJB equation. Minimization over the forward-drift control field then
determines the optimal feedback drift. The corresponding local HJB equation,
coupled to the density-state equation, reduces to the Madelung equations and,
after imposing the standard single-valuedness condition on the reconstructed
phase, to the Schr\"odinger equation. Under the Madelung
reconstruction, the stochastic process describes a definite but random electron
position with
probability density \(P=|\psi|^2\), so its one-time position law is the Born
position law.

For the hydrogen atom, the formulation gives an explicit optimal feedback drift
for each stationary eigenstate \(\psi_{n\ell m}\), whose It\^o equations in
spherical coordinates can be integrated directly. To our knowledge, the
resulting simulations are the first at trajectory level for hydrogen
eigenstates with nonzero angular momentum, and the first in which the
electron's energies and angular momentum are recovered from the
trajectories themselves. The simulated radial, polar,
and azimuthal coordinate distributions converge to the corresponding Born
marginals after sufficient statistical averaging. The same drift fields
reproduce the standard radial and angular kinetic-energy expectation values,
and the computed total energies agree with the hydrogen energy spectrum for
the states examined. For the magnetic eigenstates \((2,1,\pm1)\) the drift
acquires a phase-driven azimuthal component: the simulated trajectories
circulate at the analytically predicted rate while sharing identical Born
marginals, and the angular momentum accumulated from the raw trajectory
increments converges to exactly \(L_z=m\hbar\), so the angular-momentum
quantization postulated in the Bohr model reappears as a deterministic
property of the stochastic motion.

The main limitation of the present nonrelativistic implementation appears near
nodal sets of \(\psi_{n\ell m}\), where the drift field becomes singular. This
produces heavy-tailed kinetic-energy fluctuations and slow convergence of some
averages. The stability criteria used here allow the relevant distributions and
averages to be recovered numerically, but the need for an effective high-velocity
cutoff indicates that a fully relativistic extension of the density-space
stochastic optimal-control formulation is required.

\appendix
\section{Derivation of the Functional Hamilton--Jacobi--Bellman Equation on Density Space}
\label{app:functional_hjb_derivation}

This appendix derives the functional HJB equation used in
Section~\ref{sec:soc-fp-foundations}. We assume sufficient
regularity, \(P>0\), and boundary conditions such that all integrations by
parts have vanishing boundary terms. Throughout
Appendices~\ref{app:functional_hjb_derivation}--\ref{app:quadratic_kinetic_choice},
the symbol \(m\) denotes the particle mass, as
in Section~\ref{sec:soc-fp-foundations}, not the magnetic quantum number.

In the density-space control problem, the density state evolves according to
the forward Fokker--Planck equation
\begin{equation}
\label{eq:app_forward_fp_operator}
\partial_t P
=
-\nabla\cdot(\mathbf b_+P)
+
D\Delta P.
\end{equation}
For compactness, define the Fokker--Planck operator
\begin{equation}
\label{eq:app_fp_operator}
F[P,\mathbf b_+]
=
-\nabla\cdot(\mathbf b_+P)
+
D\Delta P .
\end{equation}
The law-dependent running cost is
\begin{equation}
\label{eq:app_law_dependent_running_cost}
L(\mathbf x,\mathbf b_+,P,t)
=
\frac{m}{2}\mathbf b_{+}^{2}
-
mD\,\mathbf b_+\cdot\nabla\log P
-
V(\mathbf x,t),
\end{equation}
which is equivalent to Eq.~\eqref{eq:law_dependent_running_cost} after using
\(mD=\hbar/2\).

The full Bellman object is the density-space value functional
\begin{equation}
\label{eq:app_mean_field_value_functional}
\mathcal V(t,P)
=
\inf_{\mathbf b_+}
\left\{
\int_t^{t_f}ds\int d^3x\,
P(\mathbf x,s)L(\mathbf x,\mathbf b_+,P,s)
+
\Phi[P(\cdot,t_f)]
\right\}.
\end{equation}
Its local value derivative is denoted
\begin{equation}
\label{eq:app_local_value_derivative}
U(\mathbf x,t)
=
\frac{\delta\mathcal V}{\delta P}(t,P_t)(\mathbf x).
\end{equation}

Bellman's principle over a short time interval \(\delta t\) gives
\begin{align}
\mathcal V(t,P)
=
\inf_{\mathbf b_+}
\Bigg\{
&
\delta t\int d^3x\,
P L(\mathbf x,\mathbf b_+,P,t)
\nonumber\\
&+
\mathcal V
\left(
t+\delta t,
P+\delta t\,F[P,\mathbf b_+]
\right)
\Bigg\}.
\label{eq:app_short_time_bellman}
\end{align}
Expanding the value functional to first order in \(\delta t\),
\begin{align}
\mathcal V
\left(
t+\delta t,
P+\delta t\,F
\right)
&=
\mathcal V(t,P)
+
\delta t\,\partial_t\mathcal V(t,P)
\nonumber\\
&\quad
+
\delta t
\int d^3x\,
U(\mathbf x,t)F[P,\mathbf b_+](\mathbf x)
+
o(\delta t).
\label{eq:app_functional_taylor_expansion}
\end{align}
Substituting Eq.~\eqref{eq:app_functional_taylor_expansion} into
Eq.~\eqref{eq:app_short_time_bellman}, cancelling \(\mathcal V(t,P)\), and
dividing by \(\delta t\), gives the functional HJB equation
\begin{align}
\label{eq:app_functional_hjb}
-\partial_t\mathcal V(t,P)
=
\inf_{\mathbf b_+}
\Bigg\{
&
\int d^3x\,
P L(\mathbf x,\mathbf b_+,P,t)
\nonumber\\
&+
\int d^3x\,
U(\mathbf x,t)
\left[
-\nabla\cdot(\mathbf b_+P)
+
D\Delta P
\right]
\Bigg\}.
\end{align}
With \(D=\hbar/(2m)\), this is Eq.~\eqref{eq:functional_hjb} of the main text.

Assuming that boundary terms vanish, the Fokker--Planck terms can be integrated
by parts:
\begin{equation}
\label{eq:app_integration_by_parts_drift}
\int d^3x\,
U\left[-\nabla\cdot(\mathbf b_+P)\right]
=
\int d^3x\,
P\,\mathbf b_+\cdot\nabla U,
\end{equation}
and
\begin{equation}
\label{eq:app_integration_by_parts_diffusion}
\int d^3x\,
U\,D\Delta P
=
\int d^3x\,
P\,D\Delta U.
\end{equation}
Therefore Eq.~\eqref{eq:app_functional_hjb} becomes
\begin{align}
\label{eq:app_functional_hjb_integrated}
-\partial_t\mathcal V(t,P)
=
\inf_{\mathbf b_+}
\int d^3x\,P
\left[
L(\mathbf x,\mathbf b_+,P,t)
+
\mathbf b_+\cdot\nabla U
+
D\Delta U
\right].
\end{align}
Using Eq.~\eqref{eq:app_law_dependent_running_cost}, this is
\begin{align}
\label{eq:app_functional_hjb_integrated_running_cost}
-\partial_t\mathcal V(t,P)
=
\inf_{\mathbf b_+}
\int d^3x\,P
\left[
\frac{m}{2}\mathbf b_{+}^{2}
+
\mathbf b_+\cdot
\left(
\nabla U-mD\nabla\log P
\right)
-
V
+
D\Delta U
\right].
\end{align}
The dependence on \(\mathbf b_+\) is now pointwise. Differentiating the
integrand with respect to \(\mathbf b_+\) gives
\begin{equation}
\label{eq:app_pointwise_optimality_condition}
m\mathbf b_+
+
\nabla U
-
mD\nabla\log P
=
0.
\end{equation}
Hence the optimal forward drift is
\begin{equation}
\label{eq:app_optimal_forward_drift_U}
\mathbf b_+
=
-\frac{1}{m}\nabla U
+
D\nabla\log P.
\end{equation}
Using \(D=\hbar/(2m)\), this is Eq.~\eqref{eq:optimal_forward_drift_U} of the
main text.

It is useful to introduce the shifted local Bellman field
\begin{equation}
\label{eq:app_shifted_bellman_field}
J
=
U
-
mD\log P.
\end{equation}
Since \(mD=\hbar/2\), this is Eq.~\eqref{eq:shifted_bellman_field}. Then
\begin{equation}
\nabla J
=
\nabla U
-
mD\nabla\log P,
\end{equation}
and the optimal drift becomes
\begin{equation}
\label{eq:app_optimal_forward_drift_J}
\mathbf b_+
=
-\frac{1}{m}\nabla J.
\end{equation}

We now derive the local shifted HJB equation. After substituting the optimal
drift, the minimized density-space Hamiltonian is
\begin{equation}
\label{eq:app_minimized_hamiltonian}
-\partial_t\mathcal V(t,P)
=
\int d^3x\,P
\left[
-
V
+
D\Delta U
-
\frac{1}{2m}
\left|
\nabla U-mD\nabla\log P
\right|^2
\right].
\end{equation}
Equivalently, using \(J=U-mD\log P\),
\begin{equation}
\label{eq:app_minimized_hamiltonian_J}
-\partial_t\mathcal V(t,P)
=
\int d^3x\,P
\left[
-
V
+
D\Delta U
-
\frac{1}{2m}|\nabla J|^2
\right].
\end{equation}

The local adjoint equation is obtained formally by taking the variational
derivative of the minimized Hamiltonian with respect to \(P\), while \(U\) is
held fixed. This is the costate (adjoint) equation of the density-space control
problem: \(U=\delta\mathcal V/\delta P\) plays the role of the costate, and its
evolution is governed by the variational derivative of the minimized
Hamiltonian density at fixed costate, as in the dynamic-programming formulation
of McKean--Vlasov control~\cite{CarmonaDelarue2018,Pham2017}.

This formal step can be made explicit along the optimal flow. Differentiating
the functional HJB equation~\eqref{eq:app_functional_hjb} with respect to
\(P\) produces, besides the variational derivative of the minimized
Hamiltonian at fixed \(U\), a second contribution in which the variation of
the minimized Hamiltonian with respect to \(U\) is paired with the second
functional derivative of \(\mathcal V\). At the optimal drift this variation
is
\[
\frac{\delta}{\delta U(\mathbf y)}
\int d^3x\,P
\left[
D\Delta U
-
\frac{1}{2m}
\left|
\nabla U-mD\nabla\log P
\right|^2
\right]
=
\nabla\cdot\left(\frac{P}{m}\nabla J\right)(\mathbf y)
+
D\Delta P(\mathbf y),
\]
which is exactly the right-hand side of the optimal Fokker--Planck
equation~\eqref{eq:app_fp_optimal_J} derived below. The second contribution
therefore accounts precisely for the transport of
\(U=\delta\mathcal V/\delta P\) along the optimal density flow, so the time
derivative \(\partial_t U\) in the local equation below is to be understood as
the total derivative of \(U(\mathbf x,t)\) along this flow, and no explicit
second-derivative term remains. Introduce the
momentum field
\begin{equation}
\label{eq:app_q_definition}
\mathbf p
=
\nabla U
-
mD\nabla\log P
=
\nabla J,
\end{equation}
which is conjugate to the shifted Bellman field \(J\) and related to the
optimal drift by \(\mathbf b_+=-\mathbf p/m\).
For the term
\[
-\int d^3x\,P\frac{|\mathbf p|^2}{2m},
\]
one obtains
\begin{equation}
\label{eq:app_variational_derivative_q_term}
\frac{\delta}{\delta P}
\left[
-\int d^3x\,P\frac{|\mathbf p|^2}{2m}
\right]
=
-\frac{|\mathbf p|^2}{2m}
-
D\nabla\cdot\mathbf p
-
D\,\mathbf p\cdot\nabla\log P.
\end{equation}
Therefore
\begin{align}
\label{eq:app_local_U_equation_general}
-\partial_t U
&=
-
V
+
D\Delta U
-
\frac{1}{2m}|\nabla J|^2
-
D\Delta J
-
D\nabla J\cdot\nabla\log P.
\end{align}
Using \(U=J+mD\log P\), this becomes
\begin{equation}
\label{eq:app_local_U_equation}
-\partial_t U
=
-
V
-
\frac{1}{2m}|\nabla J|^2
+
mD^2\Delta\log P
-
D\nabla J\cdot\nabla\log P.
\end{equation}

Substituting the optimal forward drift
Eq.~\eqref{eq:app_optimal_forward_drift_J} into the
Fokker--Planck state equation~\eqref{eq:app_forward_fp_operator}
gives the corresponding closed-loop density evolution:
\begin{equation}
\label{eq:app_fp_optimal_J}
\partial_t P
=
\nabla\cdot\left(\frac{P}{m}\nabla J\right)
+
D\Delta P.
\end{equation}
Dividing by \(P\), where \(P>0\), gives
\begin{equation}
\label{eq:app_logP_time_from_J}
\partial_t\log P
=
\frac{1}{m}
\left(
\Delta J+\nabla J\cdot\nabla\log P
\right)
+
D
\left(
\Delta\log P+|\nabla\log P|^2
\right).
\end{equation}
Since
\begin{equation}
\label{eq:app_J_time_relation_U}
J=U-mD\log P,
\end{equation}
we have
\begin{equation}
\label{eq:app_minus_J_time_relation}
-\partial_t J
=
-\partial_t U
+
mD\,\partial_t\log P.
\end{equation}
Substituting Eqs.~\eqref{eq:app_local_U_equation} and
\eqref{eq:app_logP_time_from_J} into Eq.~\eqref{eq:app_minus_J_time_relation}
yields
\begin{align}
\label{eq:app_shifted_hjb_log_terms}
-\partial_t J
&=
-
V
-
\frac{1}{2m}|\nabla J|^2
+
D\Delta J
+
2mD^2\Delta\log P
+
mD^2|\nabla\log P|^2.
\end{align}
Using
\begin{equation}
\label{eq:app_sqrt_identity_for_J}
\frac{\Delta\sqrt P}{\sqrt P}
=
\frac{1}{2}\Delta\log P
+
\frac{1}{4}|\nabla\log P|^2,
\end{equation}
the last two terms combine as
\begin{equation}
2mD^2\Delta\log P
+
mD^2|\nabla\log P|^2
=
4mD^2\frac{\Delta\sqrt P}{\sqrt P}.
\end{equation}
Thus the shifted local HJB equation is
\begin{equation}
\label{eq:app_hjb_density_coupled_optimal}
-\partial_t J
=
-
V
-
\frac{1}{2m}|\nabla J|^2
+
D\Delta J
+
4mD^2\frac{\Delta\sqrt P}{\sqrt P}.
\end{equation}
Since \(D=\hbar/(2m)\), this is equivalently
\begin{equation}
\label{eq:app_hjb_density_coupled_optimal_hbar}
-\partial_t J
=
-
V
-
\frac{1}{2m}|\nabla J|^2
+
\frac{\hbar}{2m}\Delta J
+
\frac{\hbar^2}{m}
\frac{\Delta\sqrt P}{\sqrt P}.
\end{equation}
This is Eq.~\eqref{eq:hjb_density_coupled_optimal} of the main text.

\section{Algebraic Reduction of the Local HJB--Fokker--Planck System to the Quantum Hamilton--Jacobi Equation}
\label{app:qhj_reduction}

This appendix derives the Madelung Hamilton--Jacobi equation from the shifted
local HJB--Fokker--Planck system
\begin{equation}
\label{eq:appB_hjb_density_coupled_optimal}
-\partial_t J
=
-
V
-
\frac{1}{2m}|\nabla J|^2
+
D\Delta J
+
4mD^2\frac{\Delta\sqrt P}{\sqrt P},
\end{equation}
and
\begin{equation}
\label{eq:appB_fp_optimal_J}
\partial_t P
=
\nabla\cdot\left(\frac{P}{m}\nabla J\right)
+
D\Delta P.
\end{equation}
The key change of variables is
\begin{equation}
\label{eq:appB_S_def}
S
=
-
J
-
mD\log P.
\end{equation}
Using \(mD=\hbar/2\), this is Eq.~\eqref{eq:value_function_definition} of the
main text.

From Eq.~\eqref{eq:appB_S_def},
\begin{equation}
\label{eq:appB_J_in_terms_of_S}
J
=
-
S
-
mD\log P.
\end{equation}
Therefore,
\begin{equation}
\label{eq:appB_time_J}
-\partial_t J
=
\partial_t S
+
mD\,\partial_t\log P,
\end{equation}
\begin{equation}
\label{eq:appB_grad_J}
\nabla J
=
-
\nabla S
-
mD\nabla\log P,
\end{equation}
and
\begin{equation}
\label{eq:appB_lap_J}
\Delta J
=
-
\Delta S
-
mD\Delta\log P.
\end{equation}

Substituting Eqs.~\eqref{eq:appB_time_J}--\eqref{eq:appB_lap_J} into
Eq.~\eqref{eq:appB_hjb_density_coupled_optimal} gives
\begin{align}
\partial_t S
+
mD\,\partial_t\log P
&=
-
V
-
\frac{1}{2m}
\left|
\nabla S+mD\nabla\log P
\right|^2
-
D\Delta S
\nonumber\\
&\quad
-
mD^2\Delta\log P
+
4mD^2\frac{\Delta\sqrt P}{\sqrt P}
\nonumber\\
&=
-
V
-
\frac{1}{2m}|\nabla S|^2
-
D\nabla S\cdot\nabla\log P
-
\frac{mD^2}{2}|\nabla\log P|^2
\nonumber\\
&\quad
-
D\Delta S
-
mD^2\Delta\log P
+
4mD^2\frac{\Delta\sqrt P}{\sqrt P}.
\label{eq:appB_after_substitution}
\end{align}

The current velocity obtained from Eq.~\eqref{eq:value_function_definition} is
\begin{equation}
\label{eq:appB_v_from_S}
\mathbf v
=
\frac{1}{m}\nabla S.
\end{equation}
Hence the continuity equation is
\begin{equation}
\label{eq:appB_continuity_PS}
\partial_t P
+
\nabla\cdot
\left(
P\frac{\nabla S}{m}
\right)
=
0.
\end{equation}
Where \(P>0\), this gives
\begin{equation}
\label{eq:appB_logP_identity}
\partial_t\log P
=
-\frac{1}{m}
\left(
\nabla\log P\cdot\nabla S+\Delta S
\right).
\end{equation}
Therefore,
\begin{equation}
\label{eq:appB_mD_logP_identity}
mD\,\partial_t\log P
=
-
D\nabla\log P\cdot\nabla S
-
D\Delta S.
\end{equation}
Substituting Eq.~\eqref{eq:appB_mD_logP_identity} into
Eq.~\eqref{eq:appB_after_substitution}, the mixed term
\(\nabla S\cdot\nabla\log P\) and the \(\Delta S\) term cancel. This gives
\begin{equation}
\label{eq:appB_qhj_intermediate}
\partial_t S
=
-
V
-
\frac{1}{2m}|\nabla S|^2
-
\frac{mD^2}{2}|\nabla\log P|^2
-
mD^2\Delta\log P
+
4mD^2\frac{\Delta\sqrt P}{\sqrt P}.
\end{equation}

Using the identity
\begin{equation}
\label{eq:appB_sqrtP_identity}
\frac{\Delta\sqrt P}{\sqrt P}
=
\frac{1}{2}\Delta\log P
+
\frac{1}{4}|\nabla\log P|^2,
\end{equation}
we have
\begin{equation}
\label{eq:appB_density_terms_identity}
-
mD^2\Delta\log P
-
\frac{mD^2}{2}|\nabla\log P|^2
=
-
2mD^2
\frac{\Delta\sqrt P}{\sqrt P}.
\end{equation}
Therefore Eq.~\eqref{eq:appB_qhj_intermediate} becomes
\begin{align}
\partial_t S
&=
-
V
-
\frac{1}{2m}|\nabla S|^2
-
2mD^2\frac{\Delta\sqrt P}{\sqrt P}
+
4mD^2\frac{\Delta\sqrt P}{\sqrt P}
\nonumber\\
&=
-
V
-
\frac{1}{2m}|\nabla S|^2
+
2mD^2\frac{\Delta\sqrt P}{\sqrt P}.
\label{eq:appB_qhj_D_form}
\end{align}
Since \(D=\hbar/(2m)\),
\begin{equation}
\label{eq:appB_quantum_coefficient}
2mD^2
=
\frac{\hbar^2}{2m}.
\end{equation}
Thus
\begin{equation}
\label{eq:appB_qhj_final}
\partial_t S
=
-
V
-
\frac{1}{2m}|\nabla S|^2
+
\frac{\hbar^2}{2m}
\frac{\Delta\sqrt P}{\sqrt P}.
\end{equation}
This is Eq.~\eqref{eq:quantum_hj_s} of the main text. Equivalently,
\begin{equation}
\label{eq:appB_qhj_standard_form}
\partial_t S
+
\frac{1}{2m}|\nabla S|^2
+
V
-
\frac{\hbar^2}{2m}
\frac{\Delta\sqrt P}{\sqrt P}
=
0.
\end{equation}
Together with the continuity equation~\eqref{eq:continuity_ps}, this is the
Madelung system.

\section{Exchange-Symmetric Quadratic Kinetic Terms and Selection of the Bilinear Form}
\label{app:quadratic_kinetic_choice}

This appendix examines how the density-space Bellman formulation depends on
the choice of exchange-symmetric quadratic kinetic term. We first classify
local scalar kinetic terms that are homogeneous and quadratic in the forward
and backward drifts, subject to rotational invariance, forward--backward
exchange symmetry, and recovery of the classical kinetic energy. We then use
the reduction of Appendix~\ref{app:qhj_reduction} to determine which member of
the resulting one-parameter family reproduces the standard quantum
Hamilton--Jacobi equation for \(D=\hbar/(2m)\).

Consider a local scalar kinetic term that is homogeneous and quadratic in
\(\mathbf b_+\) and \(\mathbf b_-\), with constant coefficients and no
additional preferred tensors. Before imposing rotational invariance, the
general quadratic form can be written in Cartesian components as
\begin{equation}
\label{eq:appC_general_tensor_quadratic}
T
=
\frac{m}{2}
\left(
A_{ij}b_+^i b_+^j
+C_{ij}b_-^i b_-^j
+2D_{ij}b_+^i b_-^j
\right).
\end{equation}
Under a rotation \(R\), invariance for every \(R\) requires
\(R^{\mathsf T}AR=A\), \(R^{\mathsf T}CR=C\), and
\(R^{\mathsf T}DR=D\). The only constant rank-two tensor invariant under all
spatial rotations is a scalar multiple of the identity. Hence
\(A_{ij}=a\delta_{ij}\), \(C_{ij}=c\delta_{ij}\), and
\(D_{ij}=(d/2)\delta_{ij}\), and Eq.~\eqref{eq:appC_general_tensor_quadratic}
reduces to the three scalar contractions
\begin{equation}
\label{eq:appC_general_rotational_quadratic}
T
=
\frac{m}{2}
\left(
a\mathbf b_{+}^{2}
+c\mathbf b_{-}^{2}
+d\,\mathbf b_+\cdot\mathbf b_-
\right).
\end{equation}
Invariance under \(\mathbf b_+\leftrightarrow\mathbf b_-\) requires
\(a=c\). In the classical limit \(\mathbf u\to0\), one has
\(\mathbf b_+=\mathbf b_-=\mathbf v\); recovery of the classical kinetic
energy \(m\mathbf v^{2}/2\) therefore requires \(2a+d=1\). Using
\(\mathbf b_\pm=\mathbf v\pm\mathbf u\), Eq.~\eqref{eq:appC_general_rotational_quadratic}
then becomes
\begin{equation}
\label{eq:appC_kappa_classification}
T
=
\frac{m}{2}
\left[
(2a+d)\mathbf v^{2}
+(2a-d)\mathbf u^{2}
\right]
=
\frac{m}{2}
\left(
\mathbf v^{2}+\kappa\mathbf u^{2}
\right),
\qquad
\kappa=2a-d=4a-1.
\end{equation}
Thus rotational invariance and exchange symmetry do not select a unique
member; they leave the parameter \(\kappa\). For the general kinetic term in
Eq.~\eqref{eq:general_symmetric_kinetic}, the identities
\(\mathbf v=\mathbf b_+-D\nabla\log P\) and
\(\mathbf u=D\nabla\log P\) give
\[
L_\kappa
=
\frac{m}{2}\mathbf b_{+}^{2}
-
mD\,\mathbf b_+\cdot\nabla\log P
+
\frac{mD^2}{2}(1+\kappa)|\nabla\log P|^2
-
V.
\]
The additional term relative to the choice \(\kappa=-1\) is proportional
to the Fisher information. Since it does not involve the control
\(\mathbf b_+\), the pointwise optimality condition and the closed-loop
Fokker--Planck equation~\eqref{eq:appB_fp_optimal_J} are unchanged, and
only the density-dependent part of the local costate equation is modified.
Using
\[
\frac{\delta}{\delta P}
\int d^3x\,P|\nabla\log P|^2
=
-4\frac{\Delta\sqrt P}{\sqrt P},
\]
the coefficient \(4mD^2\) in the shifted local HJB
equation~\eqref{eq:appB_hjb_density_coupled_optimal} is replaced by
\(4mD^2-2mD^2(1+\kappa)=2mD^2(1-\kappa)\). Repeating the reduction of
Appendix~\ref{app:qhj_reduction} therefore gives
\[
\partial_t S
=
-
V
-
\frac{1}{2m}|\nabla S|^2
-
2\kappa mD^2
\frac{\Delta\sqrt P}{\sqrt P}.
\]
For \(D=\hbar/(2m)\), the standard quantum Hamilton--Jacobi equation is
obtained for \(\kappa=-1\).

\section{Stationary Wave Function of the Electron in the Hydrogen Atom}
\label{app:wave_function_hydrogen}
The wave function of an electron in a hydrogen atom, which is a solution to the stationary Schrödinger equation, can be separated into radial and angular parts, each depending on specific quantum numbers. This appendix provides detailed expressions for both components.
\begin{equation}
\label{eq:psi_eq_R_Y}
\psi_{n\ell m}(r, \theta, \phi) = R_{n\ell}(r) Y_{\ell m}(\theta, \phi),
\end{equation}
where $R_{n\ell}(r)$ is the radial part and $Y_{\ell m}(\theta, \phi)$ is the angular part of the wave function.

The radial part of the wave function depends on the quantum numbers $n$ (principal quantum number) and $\ell$ (orbital angular momentum quantum number):
\begin{equation}
\label{eq:radial_wave_function}
R_{n\ell}(r) = \sqrt{\left( \frac{2}{na_0} \right)^3 \frac{(n-\ell-1)!}{2n(n+\ell)!}} \, e^{-\frac{r}{na_0}} \left( \frac{2r}{na_0} \right)^\ell L_{n-\ell-1}^{2\ell+1}\left( \frac{2r}{na_0} \right),
\end{equation}
where $a_0=\frac{4 \pi \epsilon_0 \hbar^2}{m_e e^2}$ is the Bohr radius and
$L_{n-\ell-1}^{2\ell+1}$ are the associated Laguerre polynomials. The
normalization of $R_{n\ell}$ is fixed by
\begin{equation}
\label{eq:radial_normalization}
\int_0^{\infty} \left|R_{n\ell}(r)\right|^2 r^2 \, dr = 1,
\end{equation}
which combined with the orthonormality of the spherical harmonics
\begin{equation}
\label{eq:angular_normalization}
\int_0^{2\pi}\!\!\int_0^{\pi}
\left|Y_{\ell m}(\theta,\phi)\right|^2 \sin\theta\, d\theta\, d\phi = 1
\end{equation}
gives the full normalization
\(\int |\psi_{n\ell m}|^2 \, d^3 x = 1\).

The angular part of the wave function is given by the spherical harmonics, which depend on the angular momentum quantum numbers $\ell$ and $m$:

\begin{equation}
\label{eq:angular_wave_function}
Y_{\ell m}(\theta, \phi) = \Theta_{\ell m}(\theta) \Phi_m(\phi),
\end{equation}
where the polar part $\Theta_{\ell m}(\theta)$ and the azimuthal function $\Phi_m(\phi)$ are defined as follows. Equivalently,
\begin{equation}
Y_{\ell m}(\theta,\phi)=(-1)^m \sqrt{\frac{(2\ell+1)}{4\pi} \frac{(\ell-m)!}{(\ell+m)!}}\, P_{\ell}^m(\cos\theta)\, e^{i m \phi}.
\end{equation}

The polar part is:
\begin{equation}
\label{eq:normalized_legendre_poly}
\Theta_{\ell m}(\theta) = \ (-1)^m \sqrt{\frac{(2\ell+1)}{4\pi} \frac{(\ell-m)!}{(\ell+m)!}} P_{\ell}^m(\cos\theta),
\end{equation}
where $P_{\ell}^m(\cos\theta)$ are the associated Legendre polynomials.

The azimuthal function is~\cite{thomson1912,arfken2013}:
\begin{equation}
\label{eq:phi_m_complex}
\Phi_m(\phi) = e^{i m \phi}.
\end{equation}

\section{Average Radial Kinetic Energy. Operator approach}
\label{sec:avg_radial_kinetic_energy_operator}
The operator of the radial kinetic energy is:
\begin{equation}
\hat T_r = -\frac{\hbar^2}{2 m_e} \left(\frac{1}{r^2}\frac{d}{dr}\left(r^2 \frac{d}{dr}\right)\right).
\end{equation}

Average Kinetic energy is:
\begin{equation}
\label{eq:expectation_value_of_kinetic_energy}
\langle T_r \rangle = \int_0^\infty r^2  R_{n\ell}^*(r)  \hat T_r R_{n\ell}(r) \, dr.
\end{equation}
\begin{equation}
\langle T_r \rangle = \int_0^\infty  r^2 R_{n\ell}^*(r) \left( -\frac{\hbar^2}{2 m_e} \frac{1}{r^2} \frac{d}{dr} \left( r^2 \frac{d}{dr} \right) \right) R_{n\ell}(r) \, dr.
\end{equation}

Given that  $R_{n\ell}(r)$  is a real function, we can drop the complex conjugate, so we have:

\begin{equation}
\label{eq:radial_KE}
\langle T_r \rangle = -\frac{\hbar^2}{2 m_e} \int_0^\infty R_{n\ell}(r) \frac{d}{dr} \left( r^2 \frac{d}{dr} R_{n\ell}(r) \right)\, dr.
\end{equation}

\section{Average Angular Kinetic Energy. Operator approach}
\label{sec:avg_angular_kinetic_energy_operator}
The angular kinetic energy is related to the angular part of the Laplacian operator, which, in spherical coordinates, is associated with the angular momentum $L^2$. The angular kinetic energy operator can be written as:

\begin{equation}
\hat{T}_{\text{angular}} = \frac{\hat L^2}{2 m_e r^2}.
\end{equation}

The expectation value of the angular kinetic energy is computed by integrating over the angular part of the wave function and applying the angular kinetic energy operator:

\begin{equation}
\langle T_{\text{angular}} \rangle = \int_0^{\infty}\!\! \int_0^{\pi}\!\! \int_0^{2\pi}  \psi^*_{n\ell m}\, \hat{T}_{\text{angular}}\, \psi_{n\ell m}\, r^2 \sin \theta \, d\phi \, d\theta \, dr,
\end{equation}

where $\psi_{n\ell m} = R_{n\ell} Y_{\ell m}$ as in Eq.~\eqref{eq:psi_eq_R_Y}.
Since  $\hat L^2 Y_{\ell m}(\theta, \phi) = \ell(\ell+1) \hbar^2 Y_{\ell m}(\theta, \phi)$, we can write:

\begin{equation}
\label{eq:expectation_value_of_angular_kinetic_energy}
\langle T_{\text{angular}} \rangle = \frac{\ell(\ell+1) \hbar^2}{2 m_e} \int_0^{\infty} r^2 R^2_{n\ell}(r) \frac{1}{r^2} dr  \int_0^{2\pi} \int_0^\pi |Y_{\ell m}(\theta, \phi)|^2 \sin \theta \, d\theta \, d\phi.
\end{equation}

Spherical harmonics are orthogonal over the angular domain:

\begin{equation}
\int_0^{2\pi} \int_0^\pi Y^*_{\ell m}(\theta, \phi) Y_{\ell' m'}(\theta, \phi) \sin \theta \, d\theta \, d\phi = \delta_{\ell\ell'} \delta_{mm'}.
\end{equation}

This means:

\begin{equation}
\int_0^{2\pi} \int_0^\pi |Y_{\ell m}(\theta, \phi)|^2 \sin \theta \, d\theta \, d\phi = 1.
\end{equation}

Given the orthogonality property, the angular kinetic energy becomes:

\begin{equation}
\label{eq:avg_angular_energy_operator}
\langle T_{\text{angular}} \rangle = \frac{\ell(\ell+1) \hbar^2}{2 m_e} \left\langle \frac{1}{r^2} \right\rangle.
\end{equation}
\section{Analytical Derivation of Kinetic-Energy Probability Distributions for the $(2,1,0)$ and $(2,0,0)$ States}
\label{app:analytical_kinetic_energy_distributions}

This appendix derives the analytical probability distributions of selected
kinetic-energy random variables for the two nodal hydrogen states treated in
the main text. For the \((2,1,0)\) state, the relevant variable is the polar
kinetic energy \(T_\theta\), associated with the drift divergence at the nodal
plane \(\theta=\pi/2\). For the \((2,0,0)\) state, the relevant variable is the
radial kinetic energy \(T_r\), associated with the drift divergence at the
radial node \(r=2a_0\).

\subsection{Distribution of \(T_\theta\) for the $(2,1,0)$ state}
\label{subapp:Ttheta_distribution_2p0}

To determine the polar kinetic-energy distribution, we start from
Eqs.~\eqref{eq:drift_velocity_spherical} and
\eqref{eq:normalized_legendre_poly}, using
\(P_1^0(\cos\theta)=\cos\theta\). For the \((2,1,0)\) state, the polar drift is
\begin{equation}
\label{eq:polar_velocity_2p0}
b_\theta
=
-\frac{\hbar}{m_e r}\tan\theta .
\end{equation}
From Eqs.~\eqref{eq:polar_and_azimuthal_kinetic_energies}
and~\eqref{eq:polar_velocity_2p0}, the corresponding polar kinetic-energy
variable is
\begin{equation}
\label{eq:polar_energy_2p0}
T_\theta
=
\frac{\hbar^2}{2m_e r^2}\tan^2\theta .
\end{equation}

We first treat \(r\) as a constant and consider \(\theta\) as the random
variable. Using Eqs.~\eqref{eq:probability_distributions}
and~\eqref{eq:normalized_legendre_poly}, the probability density of \(\theta\)
for the \((2,1,0)\) state is
\begin{equation}
\label{eq:polar_density_2p0}
P_\theta(\theta)
=
\frac{3}{2}\sin\theta\cos^2\theta .
\end{equation}

We use the standard transformation formula for probability densities:
\begin{equation}
\label{eq:pdf_transformation}
P_{T_\theta}(T_\theta)
=
\sum_{\theta_i:\,T_\theta(\theta_i)=T_\theta}
P_\theta(\theta_i)
\left|
\frac{d\theta_i}{dT_\theta}
\right| ,
\end{equation}
where the sum is over the branches of the inverse function
\(\theta(T_\theta)\).

From Eq.~\eqref{eq:polar_energy_2p0}, the two branches are
\begin{equation}
\theta_1(T_\theta)
=
\arctan
\sqrt{\frac{2m_e r^2}{\hbar^2}T_\theta},
\qquad
0\leq \theta_1\leq \frac{\pi}{2},
\end{equation}
and
\begin{equation}
\theta_2(T_\theta)
=
\pi-
\arctan
\sqrt{\frac{2m_e r^2}{\hbar^2}T_\theta},
\qquad
\frac{\pi}{2}\leq \theta_2\leq \pi .
\end{equation}
For both branches,
\begin{equation}
P_\theta(\theta_i)
=
\frac{3}{2}
\frac{
\sqrt{\frac{2m_e r^2}{\hbar^2}T_\theta}
}{
\left(
1+\frac{2m_e r^2}{\hbar^2}T_\theta
\right)^{3/2}
},
\end{equation}
and
\begin{equation}
\left|
\frac{d\theta_i}{dT_\theta}
\right|
=
\frac{2m_e r^2}{\hbar^2}
\frac{
1
}{
2
\sqrt{\frac{2m_e r^2}{\hbar^2}T_\theta}
\left(
1+\frac{2m_e r^2}{\hbar^2}T_\theta
\right)
}.
\end{equation}
Substituting these expressions into Eq.~\eqref{eq:pdf_transformation}, the two
branches contribute equally and give
\begin{equation}
\label{eq:lomax_polar_KE}
P(T_\theta\mid r)
=
\frac{3}{2}
\frac{2m_e r^2}{\hbar^2}
\left[
1+
\frac{2m_e r^2}{\hbar^2}T_\theta
\right]^{-5/2}.
\end{equation}
This is the exact conditional probability density of \(T_\theta\) at fixed
\(r\). It is a Lomax, or Pareto Type~II, distribution with shape parameter
\(\alpha=3/2\) and scale parameter
\begin{equation}
\lambda_\theta(r)
=
\frac{\hbar^2}{2m_e r^2}.
\end{equation}

For fixed \(r\), the large-energy regime is
\begin{equation}
\label{eq:Ttheta_large_condition}
T_\theta
\gg
\frac{\hbar^2}{2m_e r^2}.
\end{equation}
Equivalently,
\[
\frac{2m_e r^2}{\hbar^2}T_\theta\gg 1.
\]
In this regime,
\begin{equation}
1+
\frac{2m_e r^2}{\hbar^2}T_\theta
\approx
\frac{2m_e r^2}{\hbar^2}T_\theta ,
\end{equation}
and Eq.~\eqref{eq:lomax_polar_KE} becomes
\begin{equation}
\label{eq:Ttheta_tail_fixed_r}
P(T_\theta\mid r)
\sim
\frac{3}{2}
\left(
\frac{\hbar^2}{2m_e r^2}
\right)^{3/2}
T_\theta^{-5/2}.
\end{equation}

If \(r\) is treated as a random variable rather than as a constant, the
unconditional density of \(T_\theta\) is obtained by integrating over \(r\):
\begin{equation}
\label{eq:Ttheta_marginal_over_r}
P_{T_\theta}(T_\theta)
=
\int_0^\infty
P(T_\theta\mid r)
P_{r,(2,1)}(r)\,dr .
\end{equation}
The radial density for the \((2,1,0)\) state is
\begin{equation}
\label{eq:radial_density_2p0}
P_{r,(2,1)}(r)
=
\frac{r^4}{24a_0^5}e^{-r/a_0}.
\end{equation}
Using Eq.~\eqref{eq:Ttheta_tail_fixed_r} in the large-\(T_\theta\) regime gives
\begin{equation}
\label{eq:Ttheta_tail_random_r}
P_{T_\theta}(T_\theta)
\sim
\frac{3}{2}
\left(
\frac{\hbar^2}{2m_e}
\right)^{3/2}
T_\theta^{-5/2}
\int_0^\infty
r^{-3}P_{r,(2,1)}(r)\,dr .
\end{equation}
The remaining integral is finite and independent of \(T_\theta\). Thus, even
when \(r\) is treated as a random variable, the large-\(T_\theta\) behavior
remains proportional to \(T_\theta^{-5/2}\). The mean of \(T_\theta\) is finite,
whereas its variance is formally infinite.

\subsection{Distribution of \(T_r\) for the $(2,0,0)$ state}
\label{subapp:Tr_distribution_2s0}

For the \((2,0,0)\) state, the radial wave function is
\begin{equation}
\label{eq:R20_explicit}
R_{2,0}(r)
=
\frac{1}{\sqrt{2}\,a_0^{3/2}}
\left(
1-\frac{r}{2a_0}
\right)
e^{-r/(2a_0)} .
\end{equation}
It has a radial node at \(r=2a_0\). From Eq.~\eqref{eq:drift_velocity_spherical},
the radial drift is
\begin{equation}
\label{eq:radial_velocity_2s0}
b_r
=
\frac{\hbar}{m_e}
\frac{d}{dr}\log|R_{2,0}(r)|
=
\frac{\hbar}{m_e}
\left[
\frac{1}{r-2a_0}
-
\frac{1}{2a_0}
\right] .
\end{equation}
Therefore the radial kinetic-energy variable is
\begin{equation}
\label{eq:radial_energy_2s0}
T_r
=
\frac{\hbar^2}{2m_e}
\left[
\frac{1}{r-2a_0}
-
\frac{1}{2a_0}
\right]^2 .
\end{equation}

To obtain the large-\(T_r\) tail of the probability distribution, we expand near
the radial node. Let
\begin{equation}
\epsilon = r-2a_0 .
\end{equation}
Then Eq.~\eqref{eq:radial_energy_2s0} becomes
\begin{equation}
T_r
=
\frac{\hbar^2}{2m_e}
\left[
\frac{1}{\epsilon}
-
\frac{1}{2a_0}
\right]^2 .
\end{equation}
The singular term dominates when
\begin{equation}
\label{eq:epsilon_small_condition}
\left|\frac{1}{\epsilon}\right|
\gg
\frac{1}{2a_0},
\qquad
\text{or equivalently}
\qquad
|\epsilon|\ll 2a_0 .
\end{equation}
In this regime,
\begin{equation}
\label{eq:Tr_near_node}
T_r(\epsilon)
\approx
\frac{\hbar^2}{2m_e\epsilon^2}.
\end{equation}
Equivalently, the corresponding large-energy condition is
\begin{equation}
\label{eq:Tr_large_condition}
T_r
\gg
\frac{\hbar^2}{8m_ea_0^2}.
\end{equation}
Thus the following derivation gives the asymptotic large-\(T_r\) tail of the
distribution, rather than the full closed-form distribution.

From Eq.~\eqref{eq:Tr_near_node},
\begin{equation}
|\epsilon(T_r)|
=
\frac{\hbar}{\sqrt{2m_eT_r}},
\qquad
\left|
\frac{d\epsilon}{dT_r}
\right|
=
\frac{\hbar}{2\sqrt{2m_e}}T_r^{-3/2}.
\end{equation}
There are two branches, corresponding to \(\epsilon>0\) and \(\epsilon<0\).

The radial Born density for the \((2,0,0)\) state is
\begin{equation}
\label{eq:radial_density_2s0}
P_{r,(2,0)}(r)
=
r^2
\left[
R_{2,0}(r)
\right]^2
=
\frac{r^2}{2a_0^3}
\left(
1-\frac{r}{2a_0}
\right)^2
e^{-r/a_0}.
\end{equation}
Expanding around \(r=2a_0\), so that \(r^2\to 4a_0^2\) and
\(e^{-r/a_0}\to e^{-2}\), gives
\begin{equation}
\label{eq:radial_density_near_node_2s0}
P_{r,(2,0)}(r)
\approx
4a_0^2
\cdot
\frac{1}{2a_0^3}
\cdot
\frac{\epsilon^2}{(2a_0)^2}
\cdot
e^{-2}
=
\frac{\epsilon^2}{2e^2a_0^3}.
\end{equation}

Using the change-of-variables formula with the two branches,
\begin{equation}
P_{T_r}(T_r)
\approx
2
\cdot
\frac{\epsilon^2}{2e^2a_0^3}
\cdot
\frac{\hbar}{2\sqrt{2m_e}}
T_r^{-3/2}.
\end{equation}
Substituting
\begin{equation}
\epsilon^2
=
\frac{\hbar^2}{2m_eT_r}
\end{equation}
from Eq.~\eqref{eq:Tr_near_node}, we obtain
\begin{equation}
\label{eq:heavy_tail_radial_2s0}
P_{T_r}(T_r)
\approx
\frac{\hbar^3}
{4\sqrt{2}\,e^2\,m_e^{3/2}a_0^3}
T_r^{-5/2},
\qquad
T_r
\gg
\frac{\hbar^2}{8m_ea_0^2}.
\end{equation}
Therefore the radial kinetic-energy distribution has the large-energy behavior
\(P_{T_r}(T_r)\sim T_r^{-5/2}\). The mean of \(T_r\) is finite, whereas its
variance is formally infinite.

\section{Convergence of the Running Angular-Momentum Average}
\label{app:Lz_convergence}

This appendix derives the convergence law of the running time
average~\eqref{eq:Lz_running_estimator} quoted in
Section~\ref{subsec:azimuthal-current}. Substituting the azimuthal equation
of motion~\eqref{eq:final_stochastic_equation_of_motion} into
Eq.~\eqref{eq:Lz_running_estimator} and using
Eq.~\eqref{eq:Lz_quantization_trajectory} gives
\begin{equation}
\label{eq:Lz_estimator_decomposition}
\bar L_z(t)
=
m\hbar
+
\frac{1}{t}\int_0^t m_e\,\sigma\,r\sin\theta\,dW^\phi.
\end{equation}
The second term is a martingale with zero mean, so
\begin{equation}
\label{eq:Lz_unbiased}
\mathbb E\left[\bar L_z(t)\right]
=
m\hbar
\qquad
\text{for all } t.
\end{equation}
By the It\^o isometry with \(\sigma^{2}\) given by
Eq.~\eqref{eq:nelson_diffusion_coefficient}, in the stationary regime in
which the trajectory samples the Born density, the variance of the running
average is
\begin{equation}
\label{eq:Lz_variance}
\operatorname{Var}\left[\bar L_z(t)\right]
=
\mathbb E\left[\left(\bar L_z(t)-m\hbar\right)^2\right]
=
\frac{m_e^2\sigma^2\left\langle r^2\sin^2\theta\right\rangle}{t}
=
\frac{m_e\hbar\left\langle r^2\sin^2\theta\right\rangle}{t}.
\end{equation}
For the \((2,1,\pm1)\) states the Born average factorizes into elementary
integrals,
\begin{equation}
\label{eq:Lz_born_averages}
\langle r^2\rangle_{2,1}
=
30\,a_0^2,
\qquad
\langle\sin^2\theta\rangle_{1,\pm1}
=
\frac34\int_0^\pi\sin^5\!\theta\,d\theta
=
\frac45,
\end{equation}
giving \(\langle r^2\sin^2\theta\rangle_{2,1,\pm1}=24\,a_0^2\). Taking the
square root of the variance~\eqref{eq:Lz_variance} defines the standard
deviation
\begin{equation}
\label{eq:Lz_std_from_variance}
\left.\Delta\bar L_z(t)\right|_{2,1,\pm1}
=
\sqrt{\operatorname{Var}\left[\bar L_z(t)\right]}
=
\hbar\,\sqrt{\frac{24\,m_e a_0^2}{\hbar\,t}},
\end{equation}
which is the fluctuation envelope~\eqref{eq:Lz_std} quoted in the main
text.

\section*{Acknowledgments}
The author expresses sincere gratitude to Prof. Asen Pashov and Asst. Prof. Lachezar Simeonov for their insightful discussions, which made valuable contributions to the improvement of this work.

\section*{Data availability statement}
All data that support the findings of this study are generated by the openly
available simulation code~\cite{github_Yordanov2024}. The repository README
provides the parameter sets and commands needed to reproduce every figure in
this article.

\bibliographystyle{unsrtnat}
\bibliography{refs}

\end{document}